\documentclass[12pt,draftclsnofoot,onecolumn]{IEEEtran}
\usepackage[applemac]{inputenc}
\usepackage{cite}
\usepackage{amsmath,amssymb,amsfonts}
\ifCLASSINFOpdf
\else
\fi
\usepackage{epsfig}
\usepackage{amsmath}%%%Macros AMS%%%
\usepackage{amsthm}%%%Macros AMS para teoremas%%%
\usepackage{amsfonts}%%%Permite usar fuentes AMS%%%
\usepackage{amssymb}%%%Para usar simbolos AMS%%%
\usepackage{hyperref}
\hypersetup{colorlinks,% 
citecolor=black,% 
filecolor=black,% 
linkcolor=black,% 
urlcolor=black,% 
pdftex} 
\usepackage{graphicx}
\usepackage{graphics}
\usepackage{float}
\usepackage{url}
\usepackage{algorithm}
\usepackage{algpseudocode}
\usepackage{setspace}
\usepackage{subfigure}
\usepackage{balance}
\usepackage{slashbox}
\usepackage{amsmath} % for align environment
\usepackage{bm}      % for \bm macro

\usepackage{hyperref}
\usepackage{color}

\usepackage{hyperref}
\usepackage[normalem]{ulem}
\useunder{\uline}{\ulined}{}%
\hypersetup{
    colorlinks=blue,
    linkcolor=blue,
    filecolor=blue,      
    urlcolor=blue,
}

\def\mi0{\mathbf{0}}

\def\ontop#1#2{\setbox0\hbox{#2}\copy0\llap{\raise\ht0\hbox{#1}}}

\usepackage{etoolbox}
\AtBeginEnvironment{matrix}{\setlength{\arraycolsep}{1.2pt}} %With this command we control the separation between columns of the matrix environment

\makeatletter
\renewcommand*\env@matrix[1][\arraystretch]{% 
  \edef\arraystretch{#1}%
  \hskip -\arraycolsep
  \let\@ifnextchar\new@ifnextchar
  \array{*\c@MaxMatrixCols c}}
\makeatother

\begin{document}

\title{Multi-User Rate Splitting in Optical Wireless Networks}

\author{\IEEEauthorblockN{Ahmad Adnan Qidan, Khulood Alazwary, Taisir El-Gorashi, Majid Safari, Senior
Member, IEEE, Harald Haas, Fellow, IEEE,  Richard V. Penty, Fellow, IEEE,  Ian H. White, Fellow, IEEE, and Jaafar M. H. Elmirghani, Fellow, IEEE}
%\thanks{The work of Syed Jafar is supported in part by funding from NSF  CCF-1617504, NSF CNS-1731384 and ARO W911NF-16-1-0215.}
\thanks{This work has been supported  by the Engineering and Physical Sciences Research Council (EPSRC), in part by the INTERNET project under Grant EP/H040536/1, and in part by the STAR project under Grant EP/K016873/1 and in part by the TOWS project under Grant EP/S016570/1. All data are provided in full in the results section of this paper.} 
%\thanks{Jorge Plata-Chaves is with the Department of Electrical Engineering (ESAT-SCD/ SISTA), Katholieke Universiteit Leuven, B-3001 Leuven, Belgium (e-mail: jplata@esat.kuleuven.be).}         
%\thanks{Dimitris Toumpakaris is with the Department of Electrical \& Computer Engineering, University of Patras, 26500, Greece (email: dtouba@upatras.gr).} 
%\thanks{Syed A. Jafar is with the Department of Electrical Engineering and Computer Sciences, University of California, Irvine, CA, 92697 USA (e-mail: syed@uci.edu).} 
}

\markboth{SUBMITTED TO IEEE  XXX~2022}
{Khulood \MakeLowercase{\emph{et al.}}: hh}
%{MORALES-CESPEDES  \MakeLowercase{\emph{et al.}}: COGNITIVE BLIND INTERFERENCE ALIGNMENT FOR MACRO-FEMTO CELLULAR NETWORKS}




\def\be{\begin{equation}}
\def\ee{\end{equation}}

\maketitle

\begin{abstract}
Optical wireless communication (OWC) has recently received massive interest as a new technology that can support the enormous data traffic increasing on daily basis. Laser-based OWC networks can provide terabits per second (Tbps) aggregate data rates. However, the emerging OWC networks require clusters of optical transmitters to provide uniform coverage for multiple users. In this context, multi-user interference (MUI) is a crucial issue that must be managed efficiently to provide high spectral efficiency. Rate splitting (RS) is proposed as a transmission scheme to serve multiple users simultaneously by splitting the message of a given user into common and private messages, and then, each user decodes the desired message following a certain procedure. In radio frequency (RF) networks, RS provides higher spectral efficiency compared with orthogonal and non-orthogonal transmission schemes. Considering the high density of OWC networks, the performance of RS is limited by the cost of providing channel state information (CSI) at transmitters and by the noise resulting from interference cancellation. In this work, a user-grouping algorithm is proposed and used to form multiple groups, each group contains users spatially clustered. Then, an outer precoder is designed to manage inter-group interference following the methodology of blind interference alignment (BIA), which reduces the requirements of CSI at RF or optical transmitters. For intra-group interference, RS is applied within each group where the users belonging to a given group receive a unique common message on which their private messages are superimposed. Furthermore, an optimization problem is formulated to allocate the power among the private messages intended to all users such that the sum rate of the network is maximized. To relax the complexity of the optimization problem, multiple multipliers are used where an algorithm iterates to determine a sub-optimal solution. The results show the effectiveness of the proposed scheme called limited CSI-RS compared to other counterpart schemes.      
\end{abstract}

\IEEEpeerreviewmaketitle

%and that the maximum sum Degrees of Freedom are attained by the macro users.


%\begin{IEEEkeywords}
%Blind Interference Alignment, Channel State Information, Degrees of Freedom, Heterogeneous Networks, Reconfigurable Antennas
%\end{IEEEkeywords}

% \fi
\IEEEpeerreviewmaketitle

\section{Introduction}
The use of the Internet has massively increased in recent years due to  emerging technologies  such as Internet of Things (IoT), robotics, video streaming, 3D printing and virtual reality (VR), etc. This unprecedented growth in data traffic causes several challenges on current  radio frequency (RF) networks including lack of resources,  power consumption and secrecy. Therefore, researchers  in both industrial and academic communities  have investigated  the possibility to define new technologies  that can work  with RF networks to support high-user demands and relax traffic-congestion in the next generation (6G) of  wireless communications \cite{Li:14,6736752,6011734}. Optical wireless communication (OWC) has received  attention as a promising technology with the potential  to overcome the drawbacks of RF networks where the optical band offers license-free bandwidth, improved secrecy and usually optical access points (APs) provide high energy efficiency compared with RF APs. An OWC network can provide an aggregate data rate in a range  of gigabit per second (Gbps) using conventional light emitting diodes (LEDs) \cite{7072557,8240590,hhja}. Like other technologies, OWC faces  several challenges   such as  light blockage, the confined  converge area of the optical AP, and the low modulation speed of LED light sources \cite{7239528,6685754}. It is worth mentioning that increasing the number of LED-based optical APs in an indoor environment can  provide a wide coverage area and seamless user-transition. However, LEDs usually are installed primarily for illumination, and therefore, increasing the number of LEDs is limited by the recommended illumination levels in such indoor environments. Alternatively, Infrared lasers such as  vertical-cavity surface-emitting lasers  (VCSELs) can be used for data transmission  since they have high modulation speeds compared to LEDs, and they are commercially available at low prices. In an indoor environment, clusters of VCSELs can be deployed on the ceiling  to provide uniform coverage. However, the total radiated power of the VCSELs must be within eye safety regulations. It was shown that VCSEL-based OWC networks can provide up to terabits per second (Tbps) aggregate data rates \cite{mo6963803,AA19901111,owc-sp22}.

Interference management is a crucial issue in high density multiple-input and multiple-output 
(MIMO) OWC scenarios where multiple users must be served simultaneously to enhance spectral efficiency. Orthogonal techniques such  as time division multiple access (TDMA) \cite{8515272}, orthogonal frequency division multiple access (OFDMA)\cite{8361407}, space division multiple access (SDMA)\cite{9141348}, code division multiple access (CDMA)\cite{8030546}, wavelength division multiple access (WDMA)\cite{osamajaafar} can be applied to allocate resources in an orthogonal fashion among users. Despite the avoidance of interference, users might experience low quality of service due to lack of resources.  Recently,  non-orthogonal multiple access (NOMA) \cite{7342274,7572968} has been proposed for OWC to manage multi-user interference (MUI) through the power domain where power allocation is performed considering the channel quality of each user. It was shown that NOMA provides  high spectral efficiency in  OWC  compared with other orthogonal transmission schemes due to the fact that  the resources of the network can be reused among users at a given time \cite{7572968}. However, if  users have comparable   channel gains, the application of NOMA becomes a challenge. As such,  NOMA relies on perfect channel state information (CSI) at the  transmitters to give users the opportunity to decode their information correctly in such OWC scenarios. From the CSI perspective, blind interference alignment (BIA) is proposed for wireless networks \cite{GWJ11} to maximize the multiplexing gain without the need for CSI at the transmitters. In particular, the precoding matrix of each user is determined following the construction  of the BIA transmission block.  In \cite{MPGV18}, a reconfigurable optical detector composed of multiple photodiodes is derived to provide a set of linearly independent channel responses from transmitters, which are essential to apply BIA in OWC. In \cite{8636954}, BIA is applied in an OWC network with limited information on user-distribution and channel coherence time at the transmitters, and its superiority in terms of the achievable user rate is demonstrated compared with transmit precoding and orthogonal schemes. It is worth mentioning that the density of the network determines the size of the BIA transmission block. In other words, BIA suffers performance-degradation in large-scale OWC networks where the channel coherence time must be large enough to deliver such a large transmission block and the noise resulting from interference cancellation increases considerably with the number of users. In \cite{8636954,9064520,9500371}, various network topology approaches including network-cenric and user-centric perspectives are designed to relax the limitations of BIA in large-scale LED-based OWC networks where multiple elastic and/or static cells are formed, and then, BIA manages the interference within each cell regardless of the neighboring  cells formed. However, the complexity of these network topology approaches can be excessively high in the emerging OWC networks using VCSELs, where  the number of transmitters is extremely high to ensure coverage. Therefore, interference management in such VCSEL-based OWC scenarios requires further investigation.

For RF networks, a new transmission scheme referred to as rate splitting (RS)  was proposed in \cite{7470942} to treat the interference and relax the requirements of CSI at the transmitters. The methodology of RS relies on splitting  the message of a given user into  common and private messages. In general, RS allocates a fraction of the total power to the private messages intended to their corresponding users, while the remaining power is devoted to deliver the common message to all users. Basically, the common message given by a public codebook in the network is  superimposed on top of private messages, and must be decoded by all users with minimum error probability. On the other hand, each user, after decoding the common message
and removing it from the received signal using successive
interference cancellation (SIC), can decode  its private message while treating  other private messages as noise. Interestingly,  the levels of power allocated to the common and private messages in RS-based transmission are highly controlled  by the accuracy of CSI at transmitters. In \cite{6289367}, a closed-form strategy for RS  is  proposed  to serve $ K $ users in a  time correlated multiple-input single-output (MISO) broadcast channel scenario  considering imperfect CSI at transmitters. It is shown that by splitting the message of a user into two parts and treating one part as  a super common message on which the other part of the message and the messages of other users are superimposed, the achievable sum rate is higher compared with benchmarking schemes such  as zero forcing (ZF) and TDMA  at the optimal degree  of freedom (DoF). In \cite{7434643}, a novel scheme called  hierarchical rate splitting (HRS) using two-layers of RS was proposed to manage intra-group and inter-group interference in MIMO networks. 
In \cite{7805217},  a topological
RS (TRS) scheme  with weighted sum interpretation is derived in $ L $-cell MISO interference channel scenarios considering  imperfect CSI at the transmitters. In  HRS and TRS, users are arranged into groups, and multiple common messages are transmitted to  the formed groups where each common message contains the private messages of the users belonging to a certain group. The superiority of HRS and TRS was demonstrated compared to traditional RS in terms of the data rate achieved  due to the fact that  theses more advanced schemes reduce the noise resulting  from the use of SIC at each user.  In \cite{7513415}, the RS strategy was applied  in a downlink multi-user MISO system to achieve high quality of service under the minimization of power consumption.

In the context of OWC, the application of all the proposed RS schemes in the literature is not straightforward due to the high density of the emerging OWC networks compared with RF networks. In \cite{HRS}, RS and HRS schemes  were implemented in Laser-based OWC to manage multi-user interference. It was shown that the data rate achieved by RS and/or HRS  decreases considerably in large size networks due to  the requirements of CSI and noise enhancement. In \cite{Kh9500371}, an optimization problem was formulated to allocate the power among the various messages of HRS  with the aim of enhancing the sum rate of the network. However, power allocation in HRS-based OWC networks involves high complexity due to the transmission of the three different messages, i.e., inner, outer and private messages. In contrast, in this work, a novel interference management scheme using RS is proposed in an OWC network to address three key problems, namely  the high number of optical transmitters and  users, the requirements of CSI and the power allocation optimization problem. The main contributions of the paper  are  as follows: 

\begin{itemize}
\item  We formulate an optimization problem with an objective function that aims to partition all users into multiple groups based on the sum rate maximization of the users belonging to each group using RS. Interestingly, the sum rate of each group cannot be determined unless all the groups are already formed and inter-group interference is aligned. Therefore, the optimum user-grouping must be found through exhaustive search. To avoid complexity, we propose a dynamic  algorithm that arranges users spatially clustered into a specific group. In this sense, users are divided into two different sets of users, main and edge users, in accordance to  certain conditions. Then, the formation of each group starts from a unique  main user, and around that main user, some of the edge users located at the closest distance  are grouped. 

\item After the formation of multiple groups, a RS-based interference management scheme, referred to as  Limited CSI-Rate Splitting (limited CSI-RS), is derived to jointly mange inter-group interference and intra-group interference. In limited CSI-RS, an outer precoder is designed to mange the interference among the groups  by following the methodology of BIA proposed in \cite{GWJ11}. While, the users belonging to each group receive a common message superimposed on top of the private messages intended to the users of that group. Mathematical expressions are derived to explain the proposed scheme in detail. 

\item Power allocation is crucial in RS-based transmission due to the combination of different messages. Considering the application of the limited CSI-RS scheme, a fixed power is allocated to the common message intended to the users of each group, and  an optimization problem is formulated  to allocate the power among the private messages of all the users,  maximizing the overall sum rate of the network. By solving the original optimization problem, an optimum solution can be determined at high complexity. Therefore, we reformulate the optimization problem around multiple  multipliers,  which can be updated separately to guarantee the provision of sub-optimal solutions with low complexity.  
\end{itemize}
The results demonstrate that our proposed dynamic user grouping algorithm converges to the optimal solution determined through exhaustive search under fixed power allocation. Moreover, the limited CSI-RS scheme achieves high performance in terms of sum data rates, BER and energy efficiency in dense OWC networks compared with other benchmarking schemes, which include BIA and RS. Finally, the formulated power allocation problem utilizes the power budget available in the network effectively, achieving data rates considerably higher than the fixed power allocation approach.

The remainder of this paper is organized as follows. In Section \ref{sec:system}, the system model of the OWC network considered is described. The fundamentals of BIA and RS used as baseline in this work are defined in Section \ref{sec:BIARS}.  In Section \ref{sec:CSIRS}, the user-grouping approach is designed, and the limited CSI-RS schemes is derived in detail.  The formulation and analysis of power allocation for the proposed scheme are derived in Section \ref{sec:power}. Finally, Sections \ref{sec:re} and \ref{sec:con} present simulation results and provide concluding remarks. 
 
{\it Notation}.  The notations considered in this work are defined as follows. First,  matrices and vectors are denoted by the bold upper case and lower case letters,  respectively. To represent identity and zero matrices with $M\times M$ dimension, we call out to   $\mathbf{I}_M$ and $\mathbf{0}_M$ notations, respectively, while $\mathbf{0}_{M,N}$  denotes a $M\times N$ zero matrix, $[\,\,]^T$ and $[\,\,]^H$ indicate  the transpose and hermitian transpose operators, respectively. Finally, $\mathbb{E}$ is the statistical expectation, and $\mathrm{col}\{\}$ is the column operator that stacks the considered vectors in a column.
\begin{figure}[t]
\begin{center}\hspace*{0cm}
\includegraphics[width=0.6\linewidth]{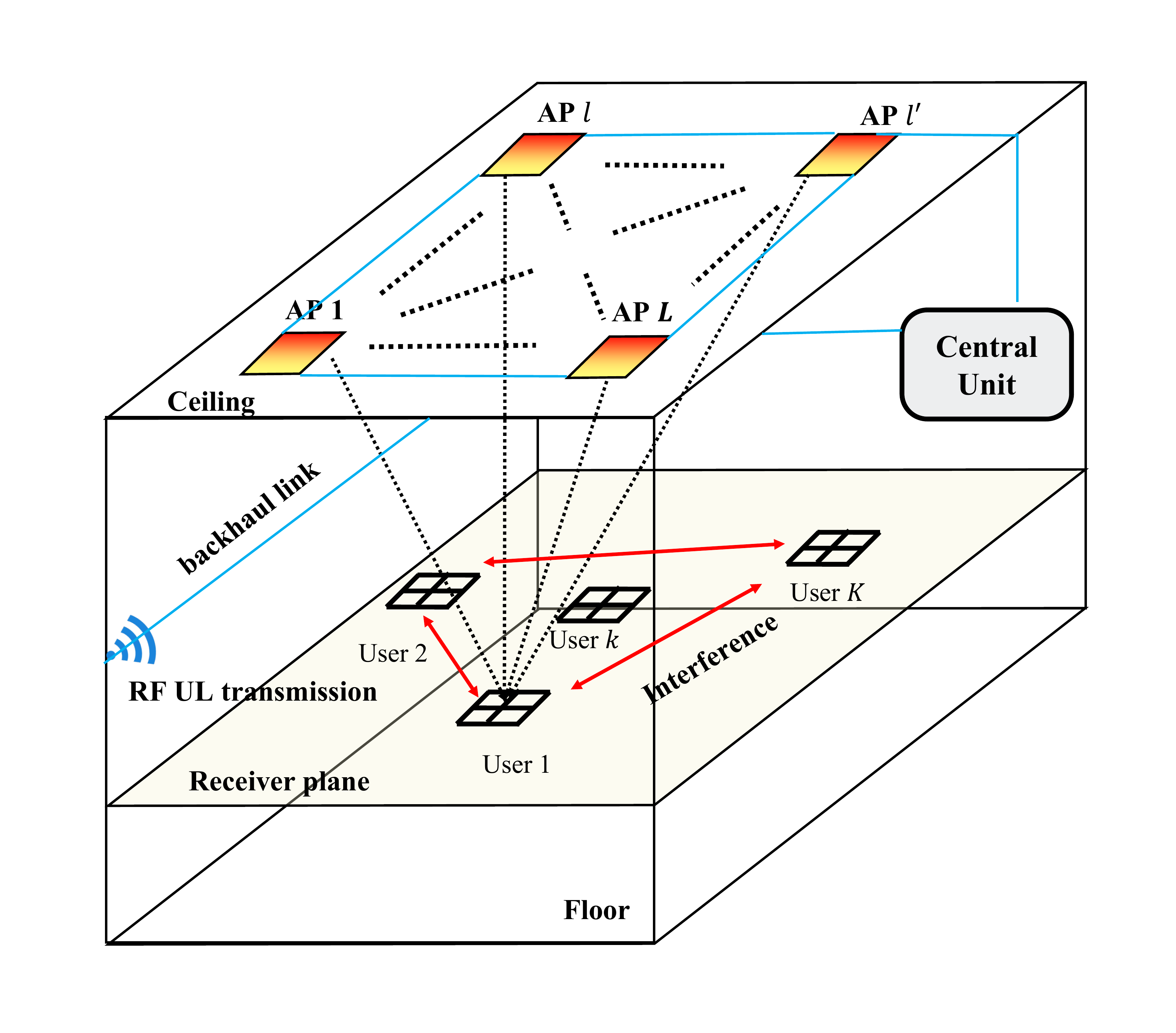}
\end{center}
\vspace{-2mm}
\caption{An OWC network composed of a number of optical APs serving multiple users.}\label{Figmodel}
\vspace{-2mm}
\end{figure}
\section{System Model}
\label{sec:system}
\begin{figure}[t]
\begin{center}\hspace*{0cm}
\includegraphics[width=0.8\linewidth]{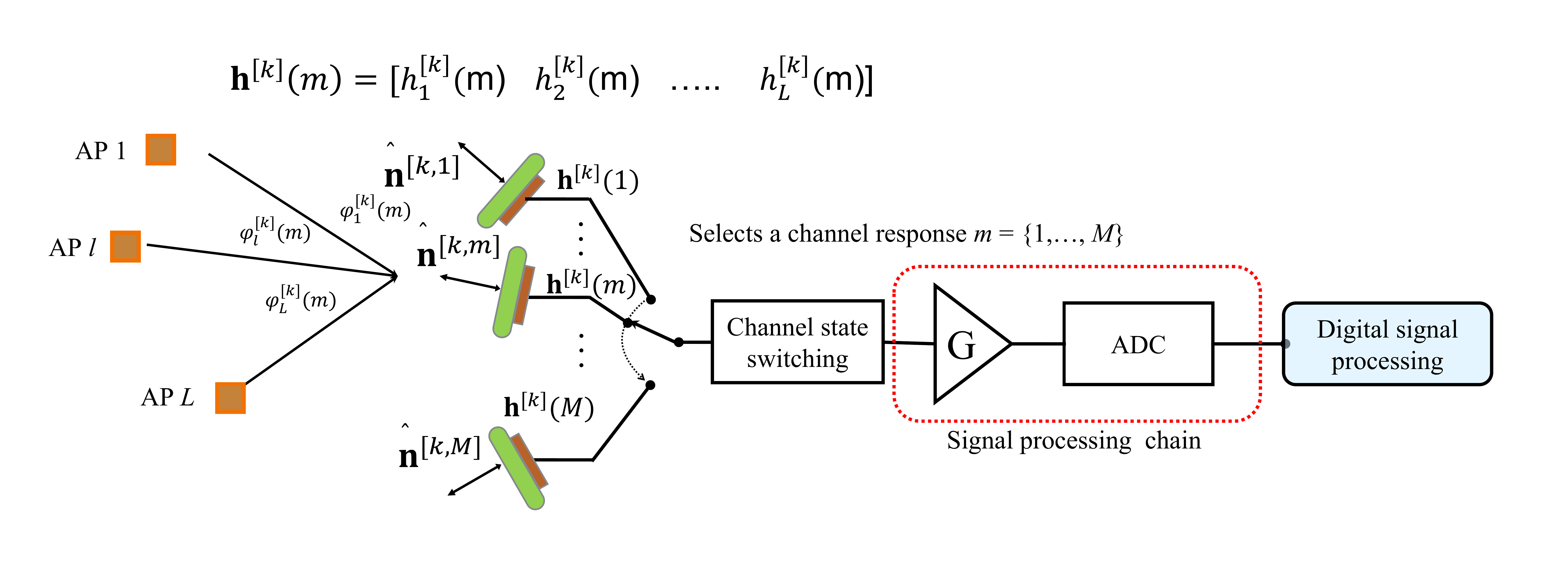}
\end{center}
\vspace{-2mm}
\caption{An optical detector with multiple photodiodes connected to a signal processing chain. }\label{Fig2}
\vspace{-2mm}
\end{figure}

An indoor downlink OWC network  is considered as  shown in Fig.~\ref{Figmodel}. It consists of multiple optical APs,  $ L, l= \big\{1, \dots, L \big\} $, installed on the ceiling to provide uniform coverage for multiple users, $ K, k= \big\{1, \dots, K \big\} $,  on the receiving plane. Each user is equipped with an optical detector  composed of multiple photodiodes, $ M, m= \big\{1, \dots, M \big\} $, providing linearly independent channel responses \cite{MPGV18,8636954}, as  shown in Fig.~\ref{Fig2}. Note that, each photodiode points to a distinct direction to guarantee connectivity, and its orientation vector  is determined by the elevation $\theta^{[k,m]}$ and azimuth $\alpha^{[k,m]}$ angles on the x-y plane, as follows 
\begin{equation}
\begin{split}
\hat{\mathbf{n}}^{[k,m]}=
\left[ \sin\left(\theta^{[k,m]} \right)\cos\left(\alpha^{[k,m]}\right), \,\, \right .
 \phantom{\{} \left . \sin\left(\theta^{[k,m]} \right)\sin\left(\alpha^{[k,m]}\right), \,\, \cos\left(\theta^{[k,m]} \right) \right],
\end{split}
\end{equation}
That is, the irradiance and incidence angles are determined by
%\begin{equation}
$\phi_{l}^{[k]}=\arccos\left(\frac{\hat{\mathbf{n}}_{l} \cdot \mathbf{v}_{l}^{[k]}}{\Vert \hat{\mathbf{n}}_{l} \Vert \Vert \mathbf{v}_{l}^{[k]} \Vert} \right)$ and
%\end{equation}
%\begin{equation}
$\varphi_{l}^{[k]}(m)=\arccos\left(\frac{\hat{\mathbf{n}}^{[k,m]} \cdot \mathbf{v}_{l}^{[k]}}{\Vert \hat{\mathbf{n}}^{[k,m]} \Vert \Vert \mathbf{v}_{l}^{[k]} \Vert}\right)$,
%\end{equation}
respectively, where $\hat{\mathbf{n}}_{l}$ is the normal vector. Considering full connectivity coordination among the $ L $ APs \footnote{Full connectivity is an AP coordination scheme that enhances the overall SNR of the network \cite{8636954}. Note that, the new RS-based scheme derived in this paper is  applicable in a multi-cell scenario. However, fractional frequency reuse must be considered to avoid inter-cell interference.}, the transmitted signal  can be written in vector form as
\begin{equation}
\mathbf{x}=\begin{bmatrix} x_{1} & x_{2} & \dots & x_{L} \end{bmatrix}^T \in \mathbb{R}_+^{L\times 1},
\end{equation}
where $ x_{l}$ is the signal transmitted by optical AP $ l $. Thus, the signal received by user $ k $  regardless of any multiple access schemes can be  written in general form  as 
\begin{equation}
y^{[k]} = \mathbf{h}^{[k]}\left(m^{[k]}\right)^{T} \mathbf{x}+z^{[k]},
\end{equation}
where $\mathbf{h}^{[k]}\left(m^{[k]}\right) \in \mathbb{R}_+^{L\times 1} $ is the channel vector between the $L$ optical APs and user $k$ at photodiode  $m$, which is given by 
\begin{equation}
 \mathbf{h}^{[k]}\left(m^{[k]}\right)= \begin{bmatrix}  h^{[k]}_{1}(m^{[k]}) & h^{[k]}_{2}(m^{[k]}) & \dots & h^{[k]}_{L}(m^{[k]}) \end{bmatrix}^{T}.
\end{equation}
Moreover, $ z^{[k]} $ is real valued additive white Gaussian noise with zero mean and variance $ \sigma^{2}_{z} $, which is given by the sum of the contributions from both shot noise and thermal noise \cite{AA19901111}.

We consider the use of VCSELs  as  transmitters under eye safety regulations \cite{mo6963803, AA19901111}.  The beam profile of the VCSEL is Gaussian dictated by the  DC bias current applied, and its 
transmitted power is determined based  on the beam waist $ W_{0} $, the wavelength $ \lambda $ and the distance $ d $ through which the beam travels. In this context, the beam radius of the VCSEL at photodiode $ m $ of user $ k $ located at the receiving plane  is given by 

\begin{equation}
W_{d}=W_{0} \left( 1+ \left(\frac{d}{d_{Ra}}\right)^{2}\right)^{1/2},
\end{equation}
where $ d_{Ra} $ is the Rayleigh range determined by $ d_{Ra}= \frac{\pi W^{2}_{0} n }{ \lambda},
 $ where $ n $ is the refractive index of the medium, which is  air in this case, i.e., $ n=1 $. In general, the spatial distribution of the intensity of  VCSEL $ l $ over the transverse plane at distance $ d $ is given by 
\begin{equation}
I_{l}(r,d) = \frac{2 P_{t,l}}{\pi W^{2}_{d}}~ \mathrm{exp}\left(-\frac{2 r^{2}}{W^{2}_{d}}\right),
\end{equation}
where $ P_{t,l} $ is the optical power. Considering that the  detection area of the optical detector  is  $ A_{rec} $, the area of each photodiode within that detector  is given by $ A_m = \frac{A_{rec}}{M} $, $ m \in M $. Hence, the  power received at each of the photodiodes  of user $ k $ from  VCSEL  $ l $ is given by 
\begin{equation}
\begin{split}
P_{m,l}=
&\int_{0}^{A_m /2 \pi} I(r,d) 2\pi r dr = P_{t,l}\left[1- \mathrm{exp}\left(- 2 \left(\frac{ A_{m}}{2 \pi W_{d}}\right)^{2}\right)\right],
\end{split}
\end{equation}
assuming photodiode $ m $ of user $ k $ is located right under transmitter $ l $, more mathematical expressions and details on the received power calculations in different scenarios can be found in \cite{mo6963803, AA19901111}.

In this work, all the optical APs are  connected through a central unit (CU) that has information regarding the distribution of the users and the channel coherence time. It also controls the resources of the network such that optimization problems with different contexts can be performed to enhance the performance of the network. Moreover, a WiFi AP is deployed to provide uplink transmission where  users can forward  the information needed for solving the optimization problems. It is worth mentioning that the need for CSI at the  transmitters in this work varies according to the transmission scheme considered to provide multiple access service in the network.

\section{ Advanced Multi-user Interference management Schemes}
\label{sec:BIARS}
In this section, we introduce the methodologies of RS and BIA in managing multi-user interference with and without CSI at transmitters, respectively, highlighting useful notations to derive the  novel scheme proposed  in Section  \ref{sec:CSIRS}.  

\subsection{Rate Splitting (RS)}
For multiple input-single output (MISO) broadcast Channel(BC) scenarios, RS can be applied to  serve $ K $ users simultaneously in  the presence of accurate CSI at transmitters \cite{7434643, HRS}. Basically,  all the transmitters  send a super common message  denoted by $s_{c} $ superimposed on the top of  $ \mathbf{s_{p}}= \begin{bmatrix} {s^{[1]}_{p}} &  {s^{[2]}_{p}} & \dots  & {s^{[K]}_{p}} \end{bmatrix}  $ private  messages intended to $ K $ users, where $ {s^{[k]}_{p}} $ is the private message desired by user $ k $. Therefore, the transmitted signal in RS-based transmission can be expressed as  
\begin{equation}
\mathbf{x}=\sqrt{P_{c}} \mathbf{w}_{c} s_{c}+\sum_{k=1}^{K} \sqrt{P_{p}} \mathbf{w}^{[k]}_{p} {s^{[k]}_{p}}
\end{equation}
where $ \mathbf{w}_{c} $ is the unit-norm precoding vector of the common message.\\  
Moreover, $  \mathbf{W_{p}}= \left[ \mathbf{w^{[1]}_{p}}  ~ \mathbf{w^{[2]}_{p}}  ~\dots ~ \mathbf{w^{[K]}_{p}} \right]   $ is a vector that contains the precoding vectors of the $ K $ private messages intended to the users  in the network. It is worth mentioning  that the power allocated to the common and private messages, $ {P_{c}} $ and $ {P_{p}} $, respectively, is usually determined through a fixed power allocation approach to avoid complexity, where  a fraction of the total power $ {P_{T}} $ is divided uniformly among the private messages, while the residual power is allocated to the common message. That is, the power allocated to the common message is $P_{c}={P_{T}}(1-t)$, where  $t \in(0,1]$, and therefore, the power allocated to each private message is  $P_{p}= \frac{{P_{T}} t}{K}$. At this point, the received signal of user $ k $ can be expressed as 
\begin{equation}
\label{cr}
{{y}}^{[k]}=\sqrt{P_{c}} {\mathbf{h}}^{[k]} \mathbf{w}_{c} {s}_{c}+ \sqrt{P_{p}} {\mathbf{h}}^{[k]} \mathbf{w}_{p}^{[k]} {s}_{p}^{[k]} \underbrace{+ \sum^{K}_{k'\neq k}\sqrt{P_{p}} {\mathbf{h}}^{[k]} \mathbf{w}_{p}^{[k']} {s}_{p}^{[k']}}_{\text{multi-user interference}}+ {{z}}^{[k]}.
\end{equation}
To decode the desired information, each user first decodes the common message with relatively low error probability, while treating all the private messages as noise. Subsequently, eliminating the common message from the received  signal through applying Successive Interference Cancellation (SIC), and then, decoding the desired private message. From \eqref{cr}, 
the SINR equations of the common and private messages can be derived  as 
\begin{equation}
\gamma^{[k]}_{c}=\frac{P_{c}\left|\mathbf{h}^{[k]^{H}} \mathbf{w}_{c}\right|^{2}}{\sum_{k=1}^{K} P_{p}\left|\mathbf{h}^{[k]^{H}} \mathbf{w}^{[k]}_{p}\right|^{2}+{\sigma_{z}}^{2}}
\end{equation}
\begin{equation}
\gamma_{p}^{[k]}=\frac{P_{p}\left|\mathbf{h}^{[k]^{H}} \mathbf{w}^{[k]}_{p}\right|^{2}}{\sum_{k' \neq k} P_{p}\left|\mathbf{h}^{[k]^{H}} \mathbf{w}^{[k']}_{p}\right|^{2}+{\sigma_{z}}^{2}}
\end{equation}
Therefore, the achievable data rate of the common message is 
$R_{c}=\log _{2}\left(1+\min\limits_{k}\left\{\gamma_{c}^{[k]}\right\}\right)$, where $\min\limits_{k}\left\{\gamma_{c}^{[k]}\right\}$ ensures that each user can decode the common message. While, the sum  rate of the private messages is 
$R_{p}=\sum_{k=1}^{K} \log _{2}\left(1+\gamma_{p}^{[k]}\right)$. As a result, the sum rate of the network considering  RS is   $R_{\text{RS}}=R_{c}+R_{p}$.

\begin{figure*}[t]
\begin{center}\hspace*{0cm}
\includegraphics[width=1\linewidth]{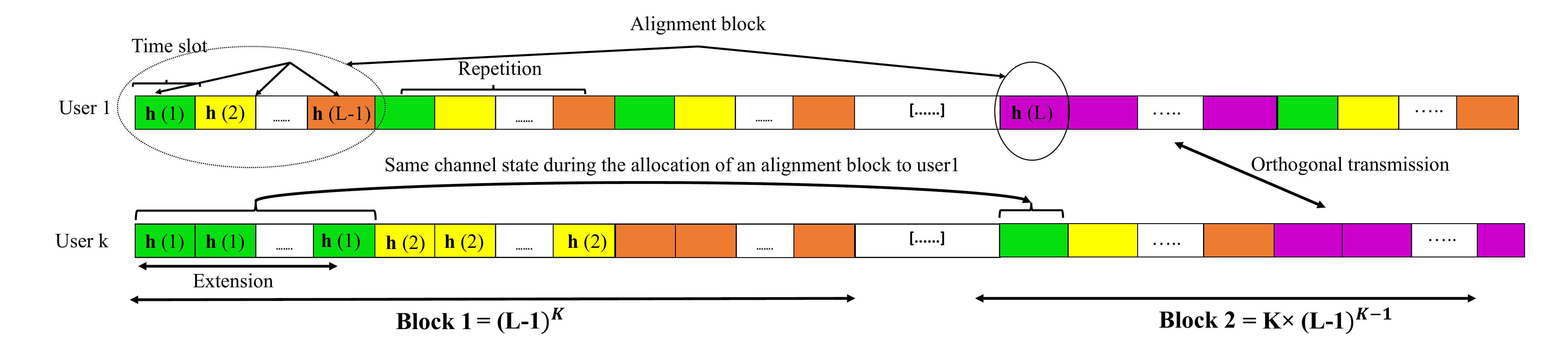}
\end{center}
\vspace{-2mm}
\caption{BIA transmission block for an OWC network. Each color represents a channel state selected by the corresponding user. }\label{Fig3}
\vspace{-2mm}
\end{figure*}

\subsection{Blind Interference Alignment (BIA)}
According to \cite{GWJ11}, BIA manages the interference among users with no CSI at transmitters  where the precoding matrix of each user is determined following the construction of a unique transmission block. In BIA, the transmission block is divided into two sub-blocks, Block 1 and  Block 2, over which users receive their information. In an OWC network with $ L $ transmitters serving $ K $ users, the whole transmission block comprises $ (L-1)^{K}+K(L-1)^{K-1} $ time slots. The first  $ (L-1)^{K} $ time slots belong to Block 1, while the last  $ K(L-1)^{K-1} $ time slots belong to Block 2, as  shown in Fig.~\ref{Fig3}. Note that, the purpose of dividing the time slots in such a way between Block 1 and Block 2 is to allocate $ \ell=\{1,  \dots, (L-1)^{K-1} \}$ orthogonal alignment blocks to each user, which determines the the precoding matrices of the users. It is worth mentioning  that each alignment block allocated to a given user $ k $ must satisfy the following two conditions:
\begin{itemize}
\item \textbf{Ensuring the decodability of each message.}  Each alignment bock consists   of $ L $ time sots to deliver the message $ \mathbf{s}_{\ell}^{[k]}= \begin{bmatrix} {s}_{\ell,1}^{[k]} & {s}_{\ell,2}^{[k]} & \dots& {s}_{\ell,L}^{[k]} \end{bmatrix}  $  from $ L $ transmitters to user $ k $, which switches its channel state over that alignment block to receive the message. 

\item \textbf{Alignment of the interference.} During transmission over the $ \ell $-th alignment block  allocated  to user $ k $, all other users, $ k' \neq k $, must remain in a fixed channel state  such that the interfering signals can be aligned in at least one dimension less than the desired information.  
\end{itemize}
These conditions can be guaranteed by allocating the first $ L-1 $ time slots of each alignment block over Block 1, while the last time slot is provided over Block 2. In other words, all users receive their information simultaneously during $ (L-1)^{K} $ time slots causing serve interference among them, which can be measured over Block 2 where users are served in an orthogonal fashion. Therefore, each user can decode its information after interference subtraction. The achievable user rate is given by 
\begin{equation}
\label{eq:rate_BIA}
R_{\text{BIA}}^{[k]} = \frac{1}{L+K-1} \log_2\left(\mathbf{I} + P_{\rm{str}}\mathbf{H}^{[k]} {\mathbf{H}^{[k]}}^H {\mathbf{R}_z}^{-1} \right),
\end{equation}
where $P_{\rm{str}}$ is the power allocated to each stream, $\mathbf{H}^{[k]} = \begin{bmatrix} \mathbf{h}^{[k]}(1) & \dots & \mathbf{h}^{[k]}(L) \end{bmatrix}^T \in \mathbb{R}^{L\times L}$ is the channel matrix of user $k$, and $\mathbf{R}_z = \begin{bmatrix}K\mathbf{I}_{L-1} & 0 \\ 0 & 1 \end{bmatrix}$.
\section{ A Novel Rate Splitting-based  Interference Management Scheme}
\label{sec:CSIRS}
The conventional  RS scheme explained in Section III is suitable for RF networks due to the large coverage area of the RF AP, which makes the provision of CSI at RF transmitters less expensive. However, RS is subject to performance-degradation in high density OWC networks in which a high number of optical APs are deployed to expand the coverage and serve multiple active users simultaneously. This is because the provision of  CSI at optical APs becomes a challenge where the information of users as well as their data must be exchanged among the whole set of optical APs. Therefore, the imperfection of CSI must be taken into consideration in such scenarios. Besides, having a high number of users means that  significant noise enhancement might occur since each user must cancel the interference received due to serving other users. One of the solutions to enhance the performance of RS in OWC is to divide the users into multiple groups from the user centric perspective. In this context, each user is grouped with users located within a threshold distance denoted by $ d_{th} $. Afterwards, RS can be applied to mange multi-user interference within each group, forming parallel RS schemes working simultaneously in multiple groups. In this case, the users of each group are subject to inter-group interference due to the fact that the same  RS precoding matrices  are used within each group. Given this, the bandwidth can be divided by the number of groups allocating an exclusive frequency or time slot to each group. Despite the low complexity of this approach, low  data rates are achieved due to insufficient use of the resources. Motivated by the methodology of BIA in managing mulit-user interference without the need for CSI, we design a novel and general framework, namely Limited CSI-RS transmission, which smartly combines the features of both BIA and RS resulting in a robust interference management scheme. In particular, once the users are divided into multiple groups, the precoding matrix of BIA can be used as an outer  precoder that manages inter-group interference with limited CSI to the channel coherence time and the distributions of users \cite{8935164}. While, the users belonging to each group receive common and private messages to align intra-group interference  using RS. It is worth pointing out that BIA suffers from  noise enhancement  as  the number of users increases (see equation \eqref{eq:rate_BIA}), and that the length of the transmission block increases with  the size of the network, which means that the channel coherence time must be large enough to ensure the delivery of  such a large transmission block with relatively minimum errors. However, in limited CSI-RS transmission, the number of groups determines the length of the transmission block instead  of the number of users, and therefore, it is expected to decrease considerably. Furthermore, RS is subject to the need for CSI, which is relaxed in limited CSI-RS transmission due to the implementation of BIA. In this section, we first formulate an optimization problem for user-grouping and provide a practical solution using a dynamic algorithm, and then, the design of the limited CSI-RS transmission scheme  is reported in detail. 

\subsection{ User grouping algorithm }
Prior to the application of the limited CSI-RS  scheme, users are divided into a number of groups, $ G $, $ g= \{ 1, \dots, G\} $, where each group $ g $ contains $ K_{g} $ active users. 
First, users are arranged into two different sets,  a main set  $ \mathcal{K}_m  $ and an edge set $ \mathcal{K}_{ed}  $. In particular, the main set contains $  \vert \mathcal{K}_{m} \vert = K_m $ main users \footnote{Users are classified as main users under two conditions; {\it i)} higher received power from most of the APs on the ceiling, {\it ii)} the minimum distance between two main users is larger than $ d_{th} $.} located at the center of the indoor environment, while the edge set contains $  \vert \mathcal{K}_{ed} \vert = K_{ed} $ edge users located around the main users. It is worth pointing out that arranging  the users into two different sets reduces the complexity of searching for the optimal set of groups  where each main user is selected as the first element in a formed group. After that, an optimization problem can be formulated  to form $ G $  groups.  The objective function of the sum rate for two users $ i \in \mathcal{K}_m $ and $ j \in \mathcal{K}_{ed} $ is given by 
\begin{equation}
U(R)= \mathit{x^{[i,j]}} \bigg(\mathit{w_i} R^{[i]} ({P_c},{P_p} )+ \mathit{w_j} R^{[j]} ({P_c},{P_p}) \bigg), \label{op21}\\
\end{equation}
where $ \mathit{x^{[i,j]}} $ is an association variable that indicates $ \mathit{x^{[i,j]}}=1 $ if users $ i $ and $ j $ are grouped together, otherwise it is $ 0 $. Moreover, $ \mathit{w_k} $, $ k\in K $, is  the  weight that achieves balance between the overall  sum rate of the users  and the system fairness, and $ R^{[k]} ({P_c},{P_p} ) $ is the user rate as function of the power levels allocated to the common and private messages. Note that, power allocation among the messages of RS intended to  the users  can  affect  the results of grouping users. However, solving these two problems, group formation and power allocation, jointly  involves significant complexity, and  therefore, a fixed power allocation approach is considered  among the users to form multiple groups.
At this point, user grouping can be performed  through maximizing the following objective function under a set of constraints 
\begin{subequations}
\label{op2}
\begin{align} 
\max_{\bm{x}} \quad &
 \sum\limits^{ K_m}_{{i=1}} \sum\limits_{{j}\in \mathcal{K}_{ed}}  U (R) \label{op21}\\
\textrm{s.t.}  \quad  & \sum\limits_{i\in \mathcal{K}_{m}} \mathit{x^{[i,j]}}=i, ~~~~~~~~~~~~~~~~~ \forall j\in \mathcal{K}_{ed}, \label{op22}\\
 \quad  & \sum\limits_{j\in K_{ed}} \mathit{x^{[i,j]}}= K_{i}, ~~~~~~~~~~~~~~~~~ \forall i\in \mathcal{K}_{m}, \label{op23}\\ 
\quad  & \mathit{x^{[i,k]}}\in \{0,1\}, K_{m}\cup K_{ed}=K,\label{op25}
\end{align} 
\end{subequations}
where the first constraint ensures that each main user belongs to one group only, while the second constraint determines the number of the edge users $ K_{i} $ associated with each main user $ i $ forming a group $ g $ with $  i\cup K_{i}=K_{g}$ users. Note that, the user rate based on limited CSI-RS transmission cannot be determined unless the groups are already formed, as explained in detail in the following sub-section. Therefore, an exhaustive search method must be adopted to solve  the optimization problem in \eqref{op2}, finding the optimal set of groups at  high computational time \cite{9500371}. Despite the optimality of exhaustive search, a practical solution is essential and is formulated as follows:

\subsubsection{Dynamic user-grouping  algorithm}
To relax the optimization problem in \eqref{op2}, the groups can be  formed separately from the maximization of the sum rate, considering the distance among the users. In the context of RS, grouping users located close to each other, i.e., users who share the same spatial correlation matrix,   maximizes the sum rate due to the relaxation  of the CSI requirements at the transmitters. Therefore, a five-step dynamic algorithm is proposed to divide users into multiple groups, while guaranteeing that the users belonging to each group are spatially clustered: 

\begin{enumerate}
\item Each main  user, $i \in \mathcal{K}_{m}$, is grouped with the closest edge user, $j \in \mathcal{K}_{ed}$, i.e.,
\begin{equation}
j = \arg\min\limits_{j\in \mathcal{K}_{ed}} \mathrm{d}(i, j),
\end{equation}
where $\mathrm{d}(i, j)$ denotes the Euclidean distance between main user $i$ and edge user $j$. Thus, $\mathrm{d}(i,j)= \sqrt{(\check{x}_{i}-\check{x}_{j})^{2}+(\check{y}_{i}-\check{y}_{j})^{2}}$, where $ (\check{x}_{i},\check{y}_{i}) $ and $ (\check{x}_{j},\check{y}_{j}) $ are the locations of users  $i$ and  $j$, respectively. 

\item The formation of a dynamic group $ g $ starts from a random pair given by main user  $i$ and edge user $j$, i.e.,  $ i\rightarrow j $, as described in the first step. Note that, the pair  $ i\rightarrow j $ must be unique and must not belong to any other group. The centroid of group $g$ at the $\kappa$-th iteration of the proposed dynamic algorithm  is defined as $\zeta_{g}(\kappa) = \left(\check{x}_{\zeta_{g}}(\kappa), \check{y}_{\zeta_{g}}(\kappa) \right)$. It is worth mentioning  that at the first iteration, the centroid of group $ g $,  \textcolor{black}{$\zeta_{g}(\kappa = 1)$}, is given by the coordinates $(\check{x}_{i},\check{y}_{i})$ of main user $i$. At this point, group $g$ is composed of $\mathcal{K}_{g} = \{i, j\}$ with the centroid $\zeta_{g}$ determined according to the coordinates of users $ i $ and $ j $.
\item The expansion of group $ g $ can be done by including the edge users located at a maximum distance less than $ d_{th} $ from  $\zeta_{g} = \left(\check{x}_{\zeta_{g}}(\kappa), \check{y}_{\zeta_{g}}(\kappa) \right)$. That is, the condition $\mathrm{d}\left( \zeta_{g}, \forall j' \in {\mathcal{K}_{ed}} \right)\leq d_{th}$ is satisfied, \footnote{ If one edge  user $ j' $ satisfies the condition of the threshold distance with more than one centroid, i.e., with more than one initial group,  during a given iteration, edge user $ j $ is assigned randomly 
to one of them.} where  $d_{th}$ is the distance that delimits the boundaries of group g. Subsequently, 
%\item  The expansion optical cell $c$ is expanded by the merging with other pairs given by the transmitter $l'$ and the user $k'$ if and only if the users $\mathcal{V}_{\mathcal{K}_c}  = \{k, k' \}$ receive a useful signal from the optical APs $\mathcal{V}_{\mathcal{L}_c} = \{l,l'\}$. That is, $\mathrm{d}\left( \forall l \in \mathcal{V}_{\mathcal{L}_c}, k' \right)\leq d_{th}$ and $\mathrm{d}\left(l',  \forall k \in \mathcal{V}_{\mathcal{K}_c} \right) \leq d_{th}$, where $d_{th}$ is the distance that delimits the footprint of each optical cell \footnote{\textcolor{black}{The footprint of each optical AP is given by the radiation semi-angle. Assuming that all the users are distributed over a plane with a distance $h$ away from the ceiling, the footprint of each optical AP is $d_{th} = h \cdot \tan\left(\phi_{1/2}\right)$.}}. 
the center of group  $g$ is updated to 
\begin{equation}
\label{eq:centr} 
\zeta_{g}(\kappa)=\left(\dfrac{\check{x}_{\zeta_{g}}(\kappa-1)+\check{x}_{j^{'}}}{2} ,\dfrac{\check{y}_{\zeta_{g}}(\kappa-1)+\check{y}_{j^{'}}}{2}\right).
\end{equation}
The dynamic  algorithm continues in the same fashion until there are no  more edge users satisfying the condition of the threshold distance with the given group $ g $, i.e, $\mathcal{K}_{g}$ is determined.
Note that,  each group  is formed iteratively including one edge user at each iteration, while updating the centroid of group $ g $ according to \eqref{eq:centr}.

\item Once the users forming group $g$ are determined, the algorithm considers grouping other edge users that do not belong to any formed group, and can be included with the user set $ \mathcal{K}_g  $. Thus, the elements of group $ g $ is updated as 

%paired to any optical AP (see step 2) and can be served by the cell $c$. Thus, the set $\mathcal{V}_{\mathcal{K}_c}$ is updated with the users contained in the footprint of the cell $c$,
% To serve as many as users, the user set $  \mathcal{V}_{\mathcal{K}_c}$  is expanded by including unselected users with priority of closest distance i.e., 
\begin{equation}
\mathcal{K}_{g}=  \mathcal{K}_{g} \cup \{{j \in \mathcal{K}_{ed}},  \mathrm{d}(\zeta_{g}(\kappa),j) \leq d_{th}\}.
\end{equation} 
Subsequently, the centroid of group $ g $ is updated again  by averaging the locations of the new included edge users according to~\eqref{eq:centr}. 

\item The proposed algorithm ends   with arranging $ K $ users into $ G $ groups, while each group with a unique main user as well as a unique set of edge users, i.e., 

\begin{equation}
\label{uniq}
~\begin{matrix}
\mathcal{K}_{g}= i \cup \mathcal{K}_{i}, \mathcal{K}_{i} \cap \mathcal{K}_{i^{'}}=\emptyset,  \mathcal{K}_{g} \cap \mathcal{K}_{g^{'}}=\emptyset, ~\\ (i \neq i^{'}, g \neq g^{'}), 
\big\{{i,i^{'}}\big\} \in \mathcal{K}_{m}, \big\{\mathcal{K}_{g}, \mathcal{K}_{g^{'}} \big\} \in K.
\end{matrix}
\end{equation}

 \end{enumerate}

%We proposed a transmission scheme that combined the BIA and RS schemes, which is subsequently referred as BIA-Rs. The state of the art idea is to apply BIA to align the inter-cluster interference, while employing Rs in intra-cluster for the purpose of easy explanation, we first focus on a toy example and after that we derived the proposed BIA-Rs scheme for the general case.

\subsection{Limited CSI-RS transmission}
After the formation of multiple groups, the Limited CSI-RS scheme is applied to manage inter-group interference using BIA as an outer precoder among the groups to relax CSI  requirements at transmitters, while the users belonging to each group receive a common message  and private messages for intra-group interference, as follows
 
\begin{figure}[t]
\begin{center}\hspace*{0cm}
\includegraphics[width=0.7\linewidth]{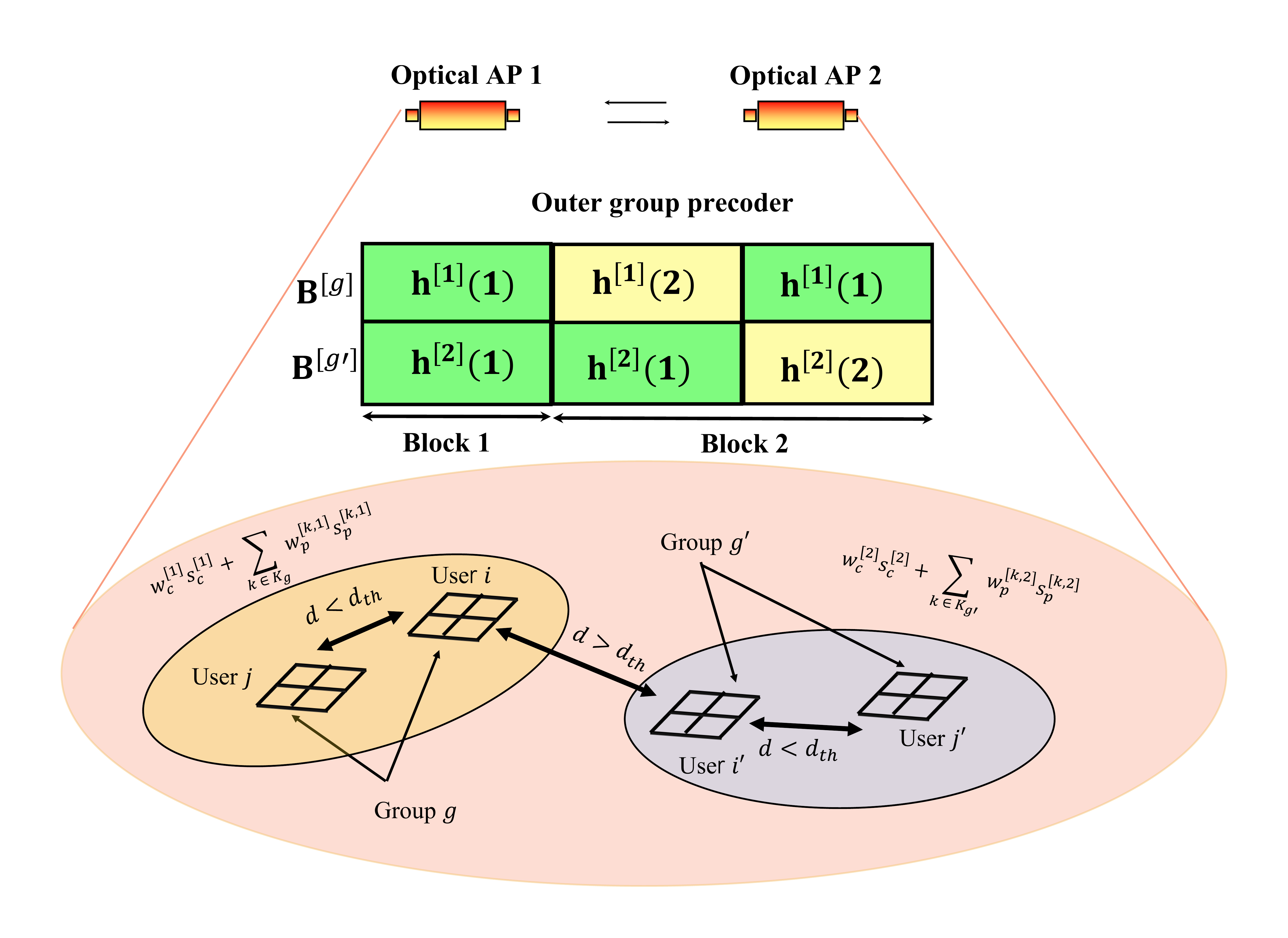}
\end{center}
\vspace{-2mm}
\caption{An OWC use case where $ L=2 $ APs serving $ G=2 $ groups, each group $ g $ with $ K_{g}=2 $ users. The groups are formed in accordance to the five steps of the dynamic algorithm. The users of each group follow the same channel state switching.}\label{Fig4}
\vspace{-2mm}
\end{figure}

\subsubsection{Motivational example}
For illustrative purposes, we first derive the mathematical expressions of the Limited CSI-RS scheme  for a toy OWC example that is composed of $ L=2 $ optical AP serving $ K=4 $ users arranged into $ G=2 $, each group with $ K_{g} $ users, as  shown in Fig.~\ref{Fig4}. In this context, the methodology of BIA explained in the previous section is followed to design the precoding matrix of each group denoted by $ \mathbf{B^{[g]}} $, i.e., the users of each group use the same outer precoding matrix, such that the interference among  the groups is aligned. Therefore, the transmitted signal considering the application of RS within each group can be expressed as 
\begin{multline}
\label{bia-rs}
\mathbf{X}= \begin{bmatrix}
\mathbf{x}[1]\\
\mathbf{x}[2]\\
\mathbf{x}[3]\\
\end{bmatrix} 
=\mathbf{B}^{[1]} \bigg(\sqrt{P_{c}^{[1]}} \mathbf{w}_{c}^{[1]} \mathbf{s}_{c}^{[1]}+\sqrt{P_{p}^{[1,1]}}\mathbf{w}_{p}^{[1,1]} \mathbf{s}_{p}^{[1,1]}+  \sqrt{P_{p}^{[2,1]}}\mathbf{w}_{p}^{[2,1]} \mathbf{s}_{p}^{[2,1]}\bigg)
\\ +\mathbf{B}^{[2]} \bigg(\sqrt{P_{c}^{[2]}} \mathbf{w}_{c}^{[2]} \mathbf{s}_{c}^{[2]}+\sqrt{P_{p}^{[1,2]}}\mathbf{w}_{p}^{[1,2]} \mathbf{s}_{p}^{[1,2]}+\sqrt{P_{p}^{[2,2]}} \mathbf{w}_{p}^{[2,2]} \mathbf{s}_{p}^{[2,2]}\bigg), 
\end{multline}
where $ {P_{c}^{[g]}} $ is the power of the common message intended to group $ g $, $ \mathbf{w}_{c}^{[g]} $   is the unit-norm precoding vector of the common
message intended to group $ g $ and $ \mathbf{s}_{c}^{[g]} $ denotes the common message intended to  users 1 and 2 belonging to group $ g $. Moreover, $ P_{p}^{[k,g]} $, $ k \in K_{g} $, is the power allocated to the private message intended to user $ k $  belonging to group $ g $, $ \mathbf{w}_{p}^{[k,g]} $ is the unit-norm precoding vector of the private message intended to user $ k \in K_{g}$, and
 $ \mathbf{s}^{[g]}_{p} =[\mathbf{s}_{p}^{[1,g]}, \mathbf{s}_{p}^{[2,g]}] \in \mathbb{C}^{2} $ is the data vector of the private messages  intended
to the user 1 and 2 of group $ g $. In this example, the transmission to two groups must occur over a predefined  transmission block comprising  three time slots in accordance to BIA methodology to manage inter-group interference as in \eqref{bia-rs}. The users of each group switch their channel pattern over  two time slots to receive information, while devoting a certain time slot to measure  the interference received due to the transmission to the adjacent group. Therefore, the BIA-based outer precoding matrices $ \mathbf{B}^{[1]} $ and $ \mathbf{B}^{[1]} $ can be expressed as      
\begin{equation}
\label{matrix}
\mathbf{B}^{[1]}= \begin{bmatrix}
\mathbf{I_2}\\
\mathbf{I_2}\\
\mathbf{0_2}\\
\end{bmatrix}, ~  \mathbf{B}^{[2]}= \begin{bmatrix}
\mathbf{I_2}\\
\mathbf{0_2}\\
\mathbf{I_2}\\
\end{bmatrix},
\end{equation}
respectively. It is worth pointing out that these outer precoding matrices are given by  0 and
1 values and CSI at transmitters is not needed to determine their structure.  Focussing on user 1 belonging to group 1, without loss of generality, the received signal can be expressed as   

\begin{multline}
\label{recived signal}
\mathbf {y}^{[1,1]}= \underbrace{\begin{bmatrix}
\mathbf{h}^{[1,1]}(1)^{T}\\ 
\mathbf{h}^{[1,1]}(2)^{T}\\
\mathbf{0}_{2,1}^{T}\\
\end{bmatrix}}_{\text{rank=2}}
\bigg(\sqrt{P_{c}^{[1]}} \mathbf{w}_{c}^{[1]} \mathbf{s}_{c}^{[1]}+ \sqrt{P_{p}^{[1,1]}}\mathbf{w}_{p}^{[1,1]} \mathbf{s}_{p}^{[1,1]}+\sqrt{P_{p}^{[2,1]}}\mathbf{w}_{p}^{[2,1]} \mathbf{s}_{p}^{[2,1]}\bigg)
+ \underbrace{\begin{bmatrix}
\mathbf{h}^{[1,1]}(1)^{T}\\
\mathbf{0}_{2,1}^{T}\\
\mathbf{h}^{[1,1]}(1)^{T} \\
\end{bmatrix}}_{\text{rank=1}}
\\\bigg(\sqrt{P_{c}^{[2]}} \mathbf{w}_{c}^{[2]} \mathbf{s}_{c}^{[2]}+ \sqrt{P_{p}^{[1,2]}}\mathbf{w}_{p}^{[1,2]} \mathbf{s}_{p}^{[1,2]}+\sqrt{P_{p}^{[2,2]}}\mathbf{w}_{p}^{[2,2]} \mathbf{s}_{p}^{[2,2]}\bigg)+\begin{bmatrix}
{z}^{[1,1]}(1)\\ 
{z}^{[1,1]}(2)\\
{z}^{1,1}(3)\\
\end{bmatrix},
\end{multline}
where $ \mathbf {y}^{[1,1]}= \text{col}\{{y}^{[1,1]}(\kappa)\}^{3}_{\kappa=1} $. It can be seen from equation \eqref{recived signal} that the desired information of user 1 appears into a full rank matrix $ [1~ 1~ 0]^{T} $, while the interference caused  by the transmission to the users belonging to group 2 is received  into one dimension $ [1~ 0~ 1]^{T} $. Thus, the desired information is decodable after measuring and canceling inter-group interference. Note that, user 1 receives  the desired information over the first time slot polluted with interference  and the second time slot, which is interference-free time slot. In other words, the first and second time slots form an alignment block over which the users of group 1 receive information. As a consequence, user 1 devotes the third time slot to measure and cancel the interference received over the first time slot of the alignment block  at the cost of noise enhancement. Thus, the received signal of user 1 belonging to group 1  after  inter-group interference cancellation can be written as  

\begin{multline}
\label{recived signal 2}
\mathbf {\widehat{y}}^{[1,1]}= {\begin{bmatrix}
\mathbf{h}^{[1,1]}(1)^{T}\\ 
\mathbf{h}^{[1,1]}(2)^{T}\\
\end{bmatrix}}
\bigg(\sqrt{P_{c}^{[1]}} \mathbf{w}_{c}^{[1]} \mathbf{s}_{c}^{[1]}+  \sqrt{P_{p}^{[1,1]}}\mathbf{w}_{p}^{[1,1]} \mathbf{s}_{p}^{[1,1]}\\+\sqrt{P_{p}^{[2,1]}}\mathbf{w}_{p}^{[2,1]} \mathbf{s}_{p}^{[2,1]}\bigg)
+\begin{bmatrix}
{z}^{[1,1]}(1)-{z}^{[1,1]}(3)\\ 
{z}^{[1,1]}(2)\\
\end{bmatrix}.
\end{multline}
It can be easily seen that the received signal of user 1 is free of inter-group interference with an increase in noise. Interestingly, $\mathbf{h}(1) $ and  $\mathbf{h}(2) $ are linearly independent channel responses, and therefore, the information of user 1 can be decoded after canceling intra-group interference by  applying the RS technique. In equation \eqref{recived signal 2}, the RS-based transmitted signal to  group 1 is  
\begin{equation}
\mathbf{X}=\bigg(\sqrt{P_{c}^{[1]}} \mathbf{w}_{c}^{[1]} \mathbf{s}_{c}^{[1]}+  \sqrt{P_{p}^{[1,1]}}\mathbf{w}_{p}^{[1,1]} \mathbf{s}_{p}^{[1,1]}+\sqrt{P_{p}^{[2,1]}}\mathbf{w}_{p}^{[2,1]} \mathbf{s}_{p}^{[2,1]}\bigg).
\end{equation}
The message of user 1 is split into common and private messages, $ \mathbf{s}_{c}^{[1]} $ and $ \mathbf{s}_{p}^{[1,1]}  $. In this case, user 1 follows the same procedure of conventional RS to decode its private message by decoding first the common message $ \mathbf{s}_{c}^{[1]} $, while treating  the private messages $ \mathbf{s}_{p}^{[1,1]}  $ and $ \mathbf{s}_{p}^{[2,1]}  $ as noise. After that, the decoded common message is  eliminated by applying SIC, giving user 1 the opportunity to decode its interference-free private message. Not that, user 2 belonging to group 1 follows the same procedure to decode the private message $ \mathbf{s}_{p}^{[2,1]}  $, as well as the other users belonging to group 2. In the following, the methodology of the limited CSI-RS transmission scheme is introduced for the general OWC scenario. 

\subsubsection{General case}
We define the limited CSI-RS transmission scheme as a general framework that is applicable for different scenarios. Thus, we consider an OWC system composed of $ L $ optical APs serving $ K $ users. The application of the proposed dynamic user-grouping algorithm arranges the $ K $ users into a set of groups $ G $, and each group contains $ K_{g} $ users. The proposed limited CSI-RS scheme has the ability to deliver interference-free signals for the users in the network  as follows:  the methodology of BIA is used to create a transmission block comprising a number of alignment blocks allocated to each group. In contrast to the conventional BIA schemes, the number of alignment blocks allocated to each group is given by  $\ell=\{1, \dots, (L-1)^{G-1}\} $, and therefore, a total of $ G\times (L-1)^{G-1} $  alignment blocks are built for all the groups over a transmission block. To satisfy  the  rule of constructing these alignment blocks while guarantee the ability to measure inter-group interference by the users belonging to each group, the transmission block must comprise $ (L-1)^{G}+ G \times(L-1)^{G-1} $ time slots. Once the construction of the transmission block is determine, the BIA-based outer precoding matrix of each group $ \mathbf{B}^{[g]} $ can be designed by arranging each alignment block $ \ell $ into a column in the precoding matrix, the rows in that column are given by $ L \times L $ identity matrices corresponding to the  time slots forming that alignment block. For instance, in equation \eqref{matrix}, the precoding matrix $  \mathbf{B}^{[1]}  $ contains 2 $ \times $ 2 identity matrices in the first and second rows corresponding to the first and second time slots that form the only alignment block allocated to group 1 given the number of transmitters and groups in that use case. It is worth mentioning that the outer precoding matrices for all the groups can be determined in this way following the structure of the BIA transmission block without the need for CSI at transmitters. Considering that the users belonging to the group use the same outer precoding matrix, the transmitted signal for the general case can be expressed as

\begin{equation}
\mathbf{X}= \sum^{G}_{g=1}\mathbf{B}^{[g]} \bigg(\sqrt{P_{c}^{[g]}} \mathbf{w}_{c}^{[g]} \mathbf{s}_{c}^{[g]}+\sqrt{P_{p}^{[k,g]}}\mathbf{W}_{p}^{[g]} \mathbf{s}_{p}^{[g]}\bigg),
\end{equation}
Note that, $ \mathbf {X}= \text{col} \{\mathbf{x}(\kappa)\}_{\kappa=1}^{\mathcal{V}} $, where $ \mathcal{V} $ is the total number of time slots that form the transmission block of BIA,   $ \mathbf{W}_{p}^{[g]}= \bigg[ \mathbf{w}_{p}^{[1,g]}, \dots, \mathbf{w}_{p}^{[k,g]}, \dots, \mathbf{w}_{p}^{[K_{g},g]} \bigg] $ is the inner precoding vector that contains  the inner precoding matrices of the private messages intended to the $ K_{g} $ users belonging to group $ g $, and $ \mathbf{s}_{c}^{[g]} =\text{col} \bigg\{\mathbf{s}_{c, \ell}^{[g]} \bigg\}_{\ell}^{(L-1)^{G-1}} $ contains the common messages transmitted to group $ g $, $ g\in G$, during the entire transmission block, where  $ \mathbf{s}_{c, \ell}^{[g]} $ is the common message transmitted over a certain alignment block $ \ell  $. Moreover, $ \mathbf{s}_{p}^{[g]}= \bigg[ \mathbf{s}_{p}^{[1,g]}, \dots, \mathbf{s}_{p}^{[k,g]}, \dots, \mathbf{s}_{p}^{[K_{g},g]} \bigg] $, where $ \mathbf{s}_{p}^{[k,g]} =\text{col} \bigg\{\mathbf{s}_{p,\ell}^{[k,g]}\bigg\}_{\ell}^{(L-1)^{G-1}} $ contains the private messages transmitted to user $ k $, $ k\in K_{g}$, over all the alignment blocks allocated to group $ g $, where $ \mathbf{s}_{p,\ell}^{[k,g]} $   is the private message transmitted over a certain alignment block $ \ell $. For detailed  mathematical explanation on the allocation of alignment blocks  to multiple groups over the BIA transmission block, we refer readers to \cite{GWJ11}. Note that, power allocation among multiple groups as well as the messages intended to the users of each group is addressed in the next section. 

The BIA-based outer precoding matrix  of each group gives the opportunity for $ K_{g} $ users  to measure and cancel  the interference received due to serving other $ K_{g'} $ users belonging to adjacent group $ g' $, $ g' \neq g $. Without loss of
generality, we consider a specific alignment block $ \ell $ allocated to group $ g $, and formulate the received signal of user $ k $ belonging to that group, $ k\in K_{g} $, after interference cancellation, which can be expressed as      
\begin{multline}
\label{eq:recgeneral}
\tilde{\mathbf{y}}^{[k,g]}=
\begin{bmatrix}
{\mathbf{h}^{[k,g]}(1)}^T\\
\vdots \\
{\mathbf{h}^{[k,g]}(L-1)}^T\\
{\mathbf{h}^{[k,g]}(L)}^T
\end{bmatrix}
\bigg(\sqrt{P_{c}^{[g]}} \mathbf{w}_{c,\ell}^{[g]} \mathbf{s}_{c,\ell}^{[g]}+\sqrt{P_{p}^{[k,g]}}\mathbf{W}_{p,\ell}^{[g]} \mathbf{s}_{p,\ell}^{[g]}\bigg)+ 
\begin{bmatrix}
z^{[k,g]}[1] - \sum\limits_{g'\neq g}^{G} z^{[k,g]}[\tau_g']\\
\vdots\\
z^{[k,g]}[L-1] - \sum\limits_{g'\neq g}^{G} z^{[k,g]}[\tau_g']\\
z^{[k,g]}[L] \\
\end{bmatrix},
\end{multline}
where $\tilde{\mathbf{y}}^{[k]} \in \mathbb{R}_+^{L\times 1}$  and the temporal index refers to the position in the alignment block rather than the corresponding temporal index within the BIA transmission block for the sake of simplicity. At this point, user $ k $ belonging to group $ g $ is subject to intra-group interference as in equation \eqref{eq:recgeneral}, which can be measured and canceled considering RS. The received signal of user $ k\in K_{g} $ polluted by intra-group interference can be rewritten as 

\begin{multline}
\tilde{\mathbf{y}}^{[k,g]}=   \sqrt{P_{c}^{[g]}} {\mathbf{H}}^{[k,g]} \mathbf{w}_{c}^{[g]} \mathbf{s}_{c}^{[g]}+ \sqrt{P_{p}^{[k,g]}} {\mathbf{H}}^{[k,g]} \mathbf{w}_{p}^{[k,g]} \mathbf{s}_{p}^{[k,g]} \underbrace{+ \sum^{K_g}_{k'\neq k}\sqrt{P_{p}^{[k',g]}} {\mathbf{H}}^{[k,g]} \mathbf{w}_{p}^{[k',g]} \mathbf{s}_{p}^{[k',g]}}_{\text{intra-group interference}}+ \tilde{\mathbf{z}}^{[k,g]}.
\end{multline}
Following the conventional RS procedure of intra-group interference cancellation, user $ k $ can decode its desired information at the cost of noise enhancement. It is worth mentioning  that  the proposed limited CSI-RS scheme results in low noise compared with traditional RS due to the fact that each user measures and cancels the private messages sent to $ K_{g} < K $ users regardless of all other users in the network.
\subsection{Achievable data rate}
We derive now the achievable sum rate of the proposed limited CSI-RS transmission scheme for the general case. Focussing on the use of RS within each group, the message of user $ k $ is divided into common and private messages. These messages are decodable, and user $ k $ belonging to group $ g $ decodes first the common message treating all the private messages as noise, and therefore, the SINR of the common message can be expressed as  
\begin{equation}
\label{snrc}
\gamma_{c}^{[k,g]}=\frac{P^{[g]}_{c}\left|\mathbf{w}^{[g]}_{c}\right|^{2}}{\sum_{k=1}^{K_g} P^{[k,g]}_{p}\left|\mathbf{w}^{[k,g]}_{p}\right|^{2}+{\sigma_{z}}^{2}}.
\end{equation}
Subsequently, user $ k $ decodes its private message treating the private messages intended to other users belonging to group $ g $  as noise. That is, the SINR of the private message is given by  
\begin{equation}
\label{snrp}
\gamma^{[k,g]}_{p}=\frac{P^{[k,g]}_{p}\left|\mathbf{w}^{[k,g]}_{p}\right|^{2}}{\sum^{K_g}_{k' \neq k} P^{[k',g]}_{p}\left| \mathbf{w}^{[k',g]}_{p}\right|^{2}+{\sigma_{z}}^{2}}.
\end{equation} 
Considering the full procedure of the limited CSI-RS scheme in managing interference among multiple groups in the network, the sum rate of the common messages is given by 
\begin{equation}
\label{ratec}
{R_{c}}=\sum_{g=1}^{G} b_{ab} \times\mathbb{E}\left[ \log \det\left( \mathbf{I} +  \gamma_{c}^{[k,g]} \mathbf{H^{[k,g]}}\mathbf{H^{[k,g]^{H}}}  \mathbf{R_{\tilde{z}}}^{-1}\right)\right],
\end{equation} 
where $  b_{ab}$ is the ratio of the alignment blocks allocated  uniformly among the groups. For $ L $ transmitters serving $ G $ groups, it is given by $ \frac{1}{G+L-1} $. Moreover, $ \mathbf{R_{\tilde{z}}} $ is 
the covariance matrix, i.e., 
\begin{equation}
\label{noisRS}
\mathbf{R_{\tilde{z}}} = 
\begin{bmatrix} 
(G-1)\mathbf{I}_{{L}-1} & \mathbf{0}\\ 
\mathbf{0} & 1\\
\end{bmatrix}.
\end{equation}
On the other hand, the sum rate of the private messages intended to $ K_g $ users belonging to each group can be expressed as 
\begin{equation}
\label{ratep}
R_{p}^{[g]}= \sum^{K_{g}}_{k=1} b_{ab} \times\mathbb{E}\left[ \log \det\left( \mathbf{I} +  \gamma_{p}^{[k,g]} \mathbf{H^{[k,g]}}\mathbf{H^{[k,g]^{H}}}  \mathbf{R_{\tilde{z}}}^{-1}\right)\right].
\end{equation} 
As a result, the overall sum rate of the proposed scheme is equal to  $ R_{\tiny{\text{limited CSI-RS}}}= R_{c}+ \sum_{g=1}^{G} R_{p}^{[g]} $.
\section{Power Allocation in limited CSI-RS transmission}
\label{sec:power}
The proposed limited CSI-RS scheme reduces the requirements of CSI at transmitters due to the formation of multiple groups guaranteeing that each group contains users spatially clustered. In addition, the outer precoding matrices of the formed groups  are deigned without CSI following the structure of the  BIA transmission block. It is worth pointing out that power allocation-based data rate maximization in the context of RS is a complex task since the power values of the common and private messages are coupled (see equations \eqref{snrc} and \eqref{snrp}). By exploiting the robustness of the proposed scheme to any imperfection in CSI at transmitters, which means intra-group interference can be relatively low, the power of the private message is defined as a vital parameter that can highly dictate the sum rate of the users. In this context, we derive a closed-form sub-optimal power allocation method that ensures  data rate maximization  compared to simply  allocating the power uniformly among the messages of the users.
\subsection{Problem formulation}
 In particular, an objective function is defined to allocate the power among the private messages intended for the users belonging to each group, while a fixed power $ P^{\dagger}_{c} $ is allocated to the common message. That is 
\begin{equation}
U (R_{\tiny{\text{limited CSI-RS}}})=  \sum^{G}_{g} \bigg( {R^{[g]}_{c}} (P^{\dagger}_{c}, P^{[k,g]}_{p})+ \sum^{K_{g}}_{k=1} {R^{[k,g]}_{p}} ( P^{[k,g]}_{p} ) \bigg),
\end{equation}
Note that, $  {R^{[g]}_{c}} (P^{\dagger}_{c},P^{[k,g]}_{p}) $ and $ {R^{[k,g]}_{p}} ( P^{[k,g]}_{p} ) $ can be easily derived from equations \eqref{ratec} and \eqref{ratep}, respectively.  In equation \eqref{uniq}, each group  is composed of a unique set of users, and therefore, the overall data rate of the network can be maximized by ensuring higher sum rate within each group. In this context, we formulate an  optimization problem  that maximizes the minimum sum rate of the $ K_g $ users  belonging to each group through controlling power allocation  among their  private messages as follows:  

\begin{equation}
\label{OP1}
\begin{aligned}
\max_{p} \quad & \bigg\{ \min_{g \in G}   {R^{[g]}_{\text{sum}}} (P^{\dagger}_{c},P^{[k,g]}_{p})  \bigg\}  \\ \\
\textrm{s.t.} \quad & {R^{[g]}_{\text{sum}}} (P^{\dagger}_{c},P^{[k,g]}_{p}) \geq R^{[g]}_{min},~~~~\\
 \quad & \sum_{k \in \mathcal{K}_{g}} P^{[k,g]}_{p} +P^{\dagger}_{c} \leq P^{[g]}_{max}, ~~~~~~~~~~\forall g \in G,\\
~~~~~~~~~~~\quad &  P^{[k,g]}_{p} \leq P^{T}_{p},~~~~~~~~~~~~~~~~~~~~~~~~~~ \forall k \in \mathcal{K}_{g},\\ 
\quad &  P^{[k,g]}_{p} \geq 0, \sum_{g\in G} P^{[g]}_{max}\leq P_{T},~~~~ k \in \mathcal{K}_{g}, g \in G,
\end{aligned}
\end{equation}
where $ {R^{[g]}_{\text{sum}}} (P^{\dagger}_{c},P^{[k,g]}_{p}) =  \bigg( {R^{[g]}_{c}} (P^{\dagger}_{c}, P^{[k,g]}_{p})+ \sum^{K_{g}}_{k=1} {R^{[k,g]}_{p}} ( P^{[k,g]}_{p} )\bigg)$. The optimization problem in \eqref{OP1} is defined as max-min fractional program under several constraints with a particular structure. It is  classified as concave-convex fractional
program that has high complexity \cite{ReP}. The first constraint guarantees that the sum rate of the users belonging to a given group $ g $ is higher than the minimum data rate $ R^{[g]}_{min} $ required to ensure high quality of service. The second constraint controls the total power allocated to the common and private messages intended to $ K_g $ users, which must be less than  or equal to the maximum power dedicated to each group $ P^{[g]}_{max} $. The third constraint limits the power allocated to the private message intended to a certain user, $ k\in K_{g} $, where it must be less than or equal to the total power $ P^{T}_{p}  $ devoted to send the private messages of the $ K_g $ users. The last constraint defines the feasible region of the optimization problem, as well as ensures that the total power consumed by $ G $ groups is less than or equal to the maximum power consumption $  P_{T} $ allowed in the network, which can be calculated according to \cite{mo6963803} under eye safety regulations due to the use of the VCSEL as an optical transmitter. 
\subsection{Sub-optimal solution}
The optimization problem in \eqref{OP1} can be solved using a parametric
approach \cite{ReP}, where it can be transformed  into  a convex
optimization problem at a given parameter value. Thus, the new  objective function can be written as
\begin{equation}
\label{CSIopt2}
f (\rho)= \max_{p}  \bigg\{ \min_{g \in G}  \Big\{ {R^{[g]}_{\text{sum}}} (P^{\dagger}_{c},P^{[k,g]}_{p})- \rho  \Big\} \bigg\},
\end{equation}
where $ \rho= P^{[g]}_{T} \delta $, considering $ P^{[g]}_{T}= P^{\dagger}_{c}  +\sum_{k \in \mathcal{K}_{g}} P^{[k,g]}_{p}  $ as  the total power  consumed by group $ g $. Moreover,  $ \delta= \Big(\min\limits_{g \in G}  \Big\{ {R^{[g]}_{\text{sum}}} (P^{\dagger}_{c},P^{[k,g]}_{p})/ P^{[g]}_{T} \Big\} \Big)$ is a non-negative parameter where at its optimal value, the power is allocated among the private messages intended to $ K_g $ users maximizing the minimum sum rate of group $ g $. According to \cite{opop}, equation \eqref{CSIopt2} can be optimally determined under the constraints in equation \eqref{OP1} through finding a
root of $f (\rho)=0$ using Dinkelbach-type algorithm. From equations \eqref{OP1} and \eqref{CSIopt2}, the optimization problem can be reformulated as

\begin{equation}
\label{OP10}
\begin{aligned}
\max_{p} \quad & \bigg\{ \min_{g \in G}  \Big\{ {R^{[g]}_{\text{sum}}} (P^{\dagger}_{c},P^{[k,g]}_{p})- \rho  \Big\} \bigg\} \\
\textrm{s.t.} \quad & \min_{g \in G}  \Big\{ {R^{[g]}_{\text{sum}}} (P^{\dagger}_{c},P^{[k,g]}_{p})- \rho  \Big\} \\ & ~~~~~~~~~~~~~~~~~~~~~~\leq {R^{[g]}_{\text{sum}}} (P^{\dagger}_{c},P^{[k,g]}_{p})- \rho,\\
\quad & {R^{[g]}_{\text{sum}}} (P^{\dagger}_{c},P^{[k,g]}_{p}) \geq R^{[g]}_{min},~~~~\\
 \quad & \sum_{k \in \mathcal{K}_{g}} P^{[k,g]}_{p} +P^{\dagger}_{c} \leq P^{[g]}_{max}, ~~~~~~~~~~\forall g \in G,\\
~~~~~~~~~~~\quad &  P^{[k,g]}_{p} \leq P^{T}_{p},~~~~~~~~~~~~~~~~~~~~~~~~~~ \forall k \in \mathcal{K}_{g},\\ 
\quad &  P^{[k,g]}_{p} \geq 0, \sum_{g\in G} P^{[g]}_{max}\leq P_{T},~~~~ k \in \mathcal{K}_{g}, g \in G.
\end{aligned}
\end{equation}
Interestingly, the optimization problem in \eqref{OP10} is a convex optimization problem due  to the fact that it has a linear  objective function under convex constraints. Therefore,  the optimal value can be equivalently found  using a distributed algorithm via Lagrangian decomposition \cite{7054502,9521837}. Given that, The
Lagrangian function of  \eqref{OP10} considering other groups in the network can be written as

\begin{equation}
\label{lag}
\mathcal{F}= \min_{g \in G}  \Big\{ {R^{[g]}_{\text{sum}}} (P^{\dagger}_{c},P^{[k,g]}_{p})- \rho  \Big\} \left(  1-\sum_{g \in G} \lambda_{g} \right) + \sum_{g \in G} \mathcal{F}_{g},
\end{equation}
where 
\begin{multline}
\label{fg}
\mathcal{F}_{g}=  \Big\{(\lambda_{g}+ \xi_{g}) {R^{[g]}_{\text{sum}}} (P^{\dagger}_{c},P^{[k,g]}_{p})- (\delta \lambda_{g}+ \nu_{g}) (P^{\dagger}_{c}+\sum_{k \in \mathcal{K}_{g}} P^{[k,g]}_{p}) \\+ \beta^{[k]}_{g} (P^{T}_{p} - P^{[k,g]}_{p}) - \xi_{g}  R^{[g]}_{min} + \nu_{g} P^{[g]}_{max}  \Big\}, 
\end{multline}
and $ \lambda_{g} $, $ \xi_{g} $, $ \nu_{g} $ and $ \beta^{[k]}_{g} $ are Lagrangian multipliers corresponding to  the four constraints in equation  \eqref{OP10}, respectively.  Therefore,
the optimal values of the power allocated to the private messages in the network can be equivalently found through maximizing $ \mathcal{F} $ in equation  \eqref{lag}. 

Using the Karush-Kuhn-Tucker (KKT) conditions \cite{100020202,2009convex}, the partial derivative $ \frac{\partial \mathcal{F}_{g} }{\partial P^{[k,g]}_{p} } $ is defined as a monotonically decreasing function with respect to the power allocated $  P^{[k,g]}_{p}  $ to the private message intended to user $ k $ belonging to group $ g $. That is, if the partial derivative $ \frac{\partial \mathcal{F}_{g} }{\partial P^{[k,g]}_{p} } \vert _{{P^{[k,g]}_{p}} =0}\leq 0 $, the optimum power value $ P^{*[k,g]}_{p}$ equals zero. Moreover, if the partial derivative $ \frac{\partial \mathcal{F}_{g} }{\partial P^{[k,g]}_{p} } \vert _{{P^{[k,g]}_{p}}=1}\geq 0 $, the optimum value $ P^{*[k,g]}_{p} $ equals one.  Otherwise, the optimum power value $ P^{*[k,g]}_{p} $ can be found as follows

\subsubsection{Optimality at  $ \beta^{[k]}_{g} $ and $ \nu_{g} $ values }
at  given values for $ \rho $, $ \lambda_{g} $ and $ \xi_{g},  $ $ \forall g \in G $, $ P^{*[k,g]}_{p} $ can be calculated by solving the following equation for each user, $ k \in K_{g} $, belonging to group $ g $ 

\begin{multline}
\label{grad2}
\frac{\partial \mathcal{F}_{g} }{\partial P^{[k,g]}_{p} }=  (\lambda_{g}+ \xi_{g}) \frac{\partial {R^{[k,g]}_{p}} (P^{[k,g]}_{p}) }{\partial P^{[k,g]}_{p} }
- (\delta \lambda_{g}+ \nu_{g})-\beta^{[k]}_{g} + \xi_{g} \frac{\partial {R^{[g]}_{c}} (P^{\dagger}_{c}, P^{[k,g]}_{p}) }{\partial P^{[k,g]}_{p} } =0.
\end{multline}
Therefore, equation \eqref{fg} can be modified taking into consideration the optimum value of the private message from \eqref{grad2} as follows 

\begin{multline}
\label{fggg}
\mathcal{F}^{*}_{g}=  \Big\{(\lambda_{g}+ \xi_{g}) {R^{[g]}_{\text{sum}}} (P^{\dagger}_{c},P^{*[k,g]}_{p})- (\delta \lambda_{g}+ \nu_{g}) (P^{\dagger}_{c}+\sum_{k \in \mathcal{K}_{g}} P^{*[k,g]}_{p})\\ + \beta^{[k]}_{g} (P^{T}_{p} - P^{*[k,g]}_{p}) - \xi_{g}  R^{[g]}_{min} + \nu_{g} P^{[g]}_{max}  \Big\}. 
\end{multline}
At this point, the gradient projection method can be applied to solve the dual problem, then, updating  the Lagrangian multipliers   $ \beta^{[k]}_{g} $ and $ \nu_{g} $  as follows 

\begin{equation}
\label{var1}
  \beta^{[k]}_{g} (\kappa)= \left[\beta^{[k]}_{g}(\kappa-1)-\epsilon_{1}(\kappa-1) \left(P^{T}_{p}-P^{[k,g]}_{p} (\kappa-1) \right) \right]^{+},
\end{equation} 
\begin{equation}
\label{var2}
\nu_{g} (\kappa)= \bigg[\nu_{g}(\kappa-1)-\epsilon_{2}(\kappa-1) \Big(P^{[g]}_{max}- \Big(P^{\dagger}_{c} +\sum_{k \in \mathcal{K}_{g}} P^{[k,g]}_{p} (\kappa-1)\Big)\Big) \bigg]^{+},
\end{equation}
where $ \kappa$ denotes the iteration of the gradient algorithm, and  $ [.]^{+} $ is a projection on the positive orthant
to take into account the fact that we have $ \beta^{[k]}_{g} , \nu_{g} \geq 0 $. Moreover, $ \epsilon_{1} $ and $ \epsilon_{2} $  are sufficient small step sizes at a given iteration $ (\kappa-1)$  that are taken in the direction of the negative gradient for the multipliers $  \beta^{[k]}_{g} $, $ \nu_{g} $, respectively. According  to equation \eqref{var2}, the Lagrangian multiplier    $   \beta^{[k]}_{g} $ works as a message to ensure that the power allocated to the private message of user $ k $ satisfies the maximum power allowed for each private message. While in equation \eqref{var2}, $ \nu_{g} $ works to ensure that the overall power allocated to the private messages intended to $ K_g $ users satisfies the maximum power constraint of group $ g $. Note that at each  iteration, the multipliers $ \beta^{[k]}_{g} $ and $ \nu_{g}  $ are updated until the optimal  value $ P^{*[k,g]}_{p}  $ is found.

\begin{algorithm}
\caption{Overall power Allocation algorithm}
\begin{algorithmic}[1]
\State {\textbf{Input}}: $ G$ and $ \rho $ ; 
\State {\textbf {Initialisation}}: $ \lambda_{g}\geq 0$, $ \xi_{g} \geq 0  $, $ \kappa=1$, $ I_{1}=1$;
\State {\text{All $ K_g $ users of each group broadcast their demands}};
\State {\text{All APs set  common values for $ \lambda_{g} $ and $ \xi_{g}  $ for each group}};
\While { $  I_{1}=1 $}
\State {\textbf {Initialisation}}: $ \beta^{[k]}_{g} \geq 0$, $ \nu_{g} \geq 0  $, $ I_{2}=1$;
\While { $  I_{2}=1 $};
\For {Each $ g\in G $};
\For {Each user $ k\in \mathcal{K}_{g}$}
\State {Allocate power according to  \eqref{grad2}};
\State {Update $ \beta^{[k]}_{g} $ according to \eqref{var1} };
\EndFor
\State {Update $ \nu_{g} $ according to \eqref{var2} };
\EndFor
\If {$ \mid P^{[k,g]}_{p} (\kappa)- P^{[k,g]}_{p} (\kappa-1)\mid\leq \epsilon $};
\State {$ I_{2}=0 $};
\Else
\State{$ {\kappa\longleftarrow \kappa+1} $};
\EndIf
\EndWhile;
\State {\textbf {Initialisation}}:$ I_{3}=1$;
\While {$  I_{3}=1 $};
\State {\text{APs determine $ \Gamma(\kappa) = \min\limits_{g} \Big\{ {R^{[g]}_{\text{sum}}} (P^{\dagger}_{c},P^{[k,g]}_{p})- \rho \Big\} $ }};
\If { $ {R^{[g]}_{\text{sum}}} (P^{\dagger}_{c},P^{[k,g]}_{p})- \rho = \Gamma(\kappa) $ or $ {R^{[g]}_{\text{sum}}} (P^{\dagger}_{c},P^{[k,g]}_{p})- \rho = \Gamma(\kappa) $ for every group with $ \lambda_{g}=0 $ };
\State {$ I_{2}=0 $};
\Else
\State{\text{Update $ \lambda_{g} $ for each group} };
\State{\text{Update power allocation according to \eqref{grad22}} };
\EndIf
\EndWhile;
\If { $ {R^{[g]}_{\text{sum}}} (P^{\dagger}_{c},P^{[k,g]}_{p})(\kappa) - {R^{[g]}_{\text{sum}}} (P^{\dagger}_{c},P^{[k,g]}_{p}) (\kappa-1)\mid\leq \epsilon $};
\State {$ I_{1}=0 $};
\Else
\State{\text{ $ \xi_{g}  $ for each group is updated according to \eqref{var4} }};
\State{$ {\kappa\longleftarrow \kappa+1} $};
\EndIf
\EndWhile
\State{\textbf{Output}}:$ P^{*[k,g]}_{p} $
\end{algorithmic}
\end{algorithm}

\subsubsection{Optimality at $ \lambda_{g} $ and $ \xi_{g}$ values }
Interestingly, the multipliers $ \lambda_{g} $ and $ \xi_{g}$ must also be updated to guarantee high quality of service for the $ K_g $ users  of each group $ g $, where at their optimal values, the power allocated to each private message determined from equation \eqref{grad2} at the optimal values of $ \beta^{*[k]}_{g} $ and $ \nu^{*}_{g}  $  must be modified to satisfy  the first and second constraints in equation \eqref{OP10}. First, at a given value of $ \xi_{g}$, the optimal value of $ \lambda_{g} $  can be determined by applying the KKT conditions. That is, 
$\frac{\partial \mathcal{F}} {\partial \Gamma } =0$, where $ \Gamma =\min\limits_{g \in G}  \Big\{ {R^{[g]}_{\text{sum}}} (P^{\dagger}_{c},P^{[k,g]}_{p})- \rho  \Big\} $. From equation \eqref{lag}, $ \sum_{g \in G} \lambda_{g} =1 $. Therefore, 
 the optimum power value $ P^{*[k,g]}_{p} $ allocated to the private message intended to each  user $ k \in K_{g} $ belonging to group $ g $  can be modified  according to
\begin{multline}
\label{grad22}
\frac{\partial \mathcal{F}_{g} }{\partial P^{[k,g]}_{p} }=  ( {\lambda}_{g}+ \widetilde{\xi}_{g}) \frac{\partial {R^{[k,g]}_{p}} (P^{[k,g]}_{p}) }{\partial P^{[k,g]}_{p} }
- (\delta \lambda_{g}+ \widetilde{\nu}^{*}_{g})-\beta^{*[k]}_{g} +\widetilde{\xi}_{g} \frac{\partial {R^{[g]}_{c}} (P^{\dagger}_{c}, P^{[k,g]}_{p}) }{\partial P^{[k,g]}_{p} } =0, 
\end{multline}
where $ \widetilde{\xi}_{g}= {\xi}_{g} \sum_{g \in G} \lambda_{g}$, $ \widetilde{\nu}_{g}={\nu}_{g} \sum_{g \in G} \lambda_{g}$ and $ \widetilde{\xi}_{g}={\xi}_{g} \sum_{g \in G} \lambda_{g}$.
Similar to equations \eqref{var1} and \eqref{var2}, the gradient projection method can be applied to obtain an updated value for the Lagrangian multiplier $ \lambda_{g} $ as follows  
\begin{multline}
\label{var3}
\lambda_{g} (\kappa)= \bigg[\lambda_{g} (\kappa-1)-\epsilon_{3}(\kappa-1) \bigg({R^{[g]}_{\text{sum}}} (P^{\dagger}_{c},P^{[k,g]}_{p}) (\kappa-1)-\rho- \min\limits_{g \in G}  \Big\{ {R^{[g]}_{\text{sum}}} (P^{\dagger}_{c},P^{[k,g]}_{p})- \rho  \Big\} \bigg) \bigg]^{+},
\end{multline}
where $ \epsilon_{3}(\kappa-1) $ is a sufficiently small step size at a given iteration $ (\kappa-1)$. Note that the optimal value $ \lambda^{*}_{g} $ determines the optimal power value allocated to the private message intended to each user for a given value $  \xi_{g} \geq 0 $. To find  the optimal value $ \lambda^{*}_{g} $,  according to the first constraint in \eqref{OP10},  $ \Big( {R^{[g]}_{\text{sum}}} (P^{\dagger}_{c},P^{[k,g]}_{p})- \rho \Big) \geq \Gamma^{*} $, where $ \Gamma^{*} = \min\limits_{g} \Big\{ {R^{[g]}_{\text{sum}}} (P^{\dagger}_{c},P^{[k,g]}_{p})- \rho \Big\} $. 
Let $ \widetilde{G} $ give the total number of the groups with $ \Big( {R^{[g]}_{\text{sum}}} (P^{\dagger}_{c},P^{[k,g]}_{p})- \rho \Big)> \Gamma^{*}$, and then, using the KKT conditions, the optimal value $ \lambda^{*}_{g} =0 $ given that 

\begin{equation}
\lambda^{*}_{g} \bigg\{ \Big( {R^{*[g]}_{\text{sum}}} (P^{\dagger}_{c},P^{[k,g]}_{p}) - \delta \Big(P^{\dagger}_{c}+ \sum\limits_{k\in K_g} P^{*[k,g]}_{p} \Big)\Big)-\Gamma^{*} \bigg\}=0.
\end{equation}
Therefore, for a given value of  $  \xi_{g} \geq 0 $, if $ \Gamma^{*} > \Big({R^{[g]}_{\text{sum}}} (P^{\dagger}_{c},P^{[k,g]}_{p})- \rho \Big)$,  $ {R^{[g]}_{\text{sum}}} (P^{\dagger}_{c},P^{[k,g]}_{p})- \rho $   cannot be less than $ \min\limits_{g} \Big\{{R^{[g]}_{\text{sum}}} (P^{\dagger}_{c},P^{[k,g]}_{p})- \rho\Big\} $.  As a consequence, the optimal value for any group $ g' $, $ g' \neq g $ not within the group set    $ \widetilde{G} $ is obtained when  $ \lambda^{*}_{g} > 0 $  and $ \Gamma^{*} = \Big({R^{[g]}_{\text{sum}}} (P^{\dagger}_{c},P^{[k,g]}_{p})- \rho \Big)$. Otherwise, $ \Big( {R^{[g]}_{\text{sum}}} (P^{\dagger}_{c},P^{[k,g]}_{p})- \rho \Big)> \Gamma^{*}$ for any group with $ \lambda^{*}_{g}=0 $.  

The optimal value of $  \xi^{*}_{g} $ must also be determined according to  
\begin{equation}
\label{var4}
\xi_{g} (\kappa)= \bigg[\xi_{g} (\kappa-1)-\epsilon_{4}(\kappa-1) \bigg({R^{[g]}_{\text{sum}}} (P^{\dagger}_{c},P^{[k,g]}_{p}) (\kappa-1)- R^{[g]}_{min} \bigg) \bigg]^{+},
\end{equation}
where $ \epsilon_{4} $ is a sufficiently small step size. It is worth pointing out that the overall algorithm iterates over the power allocated to the private message intended to each user, until the optimal value $ P^{*[k,g]}_{p}   $ is determined, which can significantly  maximize the sum rate of the $ K_g $ users belonging to each group with respect to power consumption. For better understanding, Algorithm 1 presents a summary of the overall algorithm, where $  \epsilon $ is a small tolerance value.    

\begin{table}
\centering
\caption{Simulation Parameters}
\begin{tabular}{|c|c|}
\hline
Parameter	& Value \\\hline
Transmitter Bandwidth	& 5 GHz \\\hline
Laser Wavelength  & 850 nm \\\hline
Laser beam waist & $ 5 \mu $m \\\hline
Physical area of the photodiode	&15 $\text{mm}^2$ \\\hline
%Transmitter semi-angle	& 45 deg \\\hline
Receiver FOV	& 60 deg \\\hline
Detector responsivity 	& 0.9 A/W \\\hline
%Refractive index of the filter	& 1.5 \\\hline
Gain of optical filter & 	1.0 \\\hline
Laser noise	& $-155~ dB/H$z \\\hline

%OFDM sub-carrier number                    &   64   \\\hline
%Transmitted power for Wi-Fi  AP	& 10 dBm \\\hline
%Bandwidth for Wi-Fi  AP	 &  10 MHz \\\hline
%Noise spectral density for Wi-Fi	& $ -75dBm/MH$z \\\hline
%Number of Antenna transmitters For Wi-Fi  AP.	& 2 \\\hline
\end{tabular}
\end{table}
\section{performance evaluation}
\label{sec:re}
In an indoor environment with  8m$ \times $ 8m$  \times $ 3m dimensions, $ L= 4\times 4 $ APs are deployed on the ceiling, each AP with $ L_{v} \times L_{v} $ VCSELs, to serve $ K=20 $ users  randomly distributed  on the receiving plane located at 2m distance from the ceiling. Each user is equipped with an optical detector consisting of $ M $ photodiodes that has the ability to provide a set of linearly independent channel responses  and  guarantee connectivity to almost all the APs in the room. All other simulation parameters are listed in \textbf{Table 1}.

\begin{figure}[t]
\begin{center}\hspace*{0cm}
\includegraphics[width=0.8\linewidth]{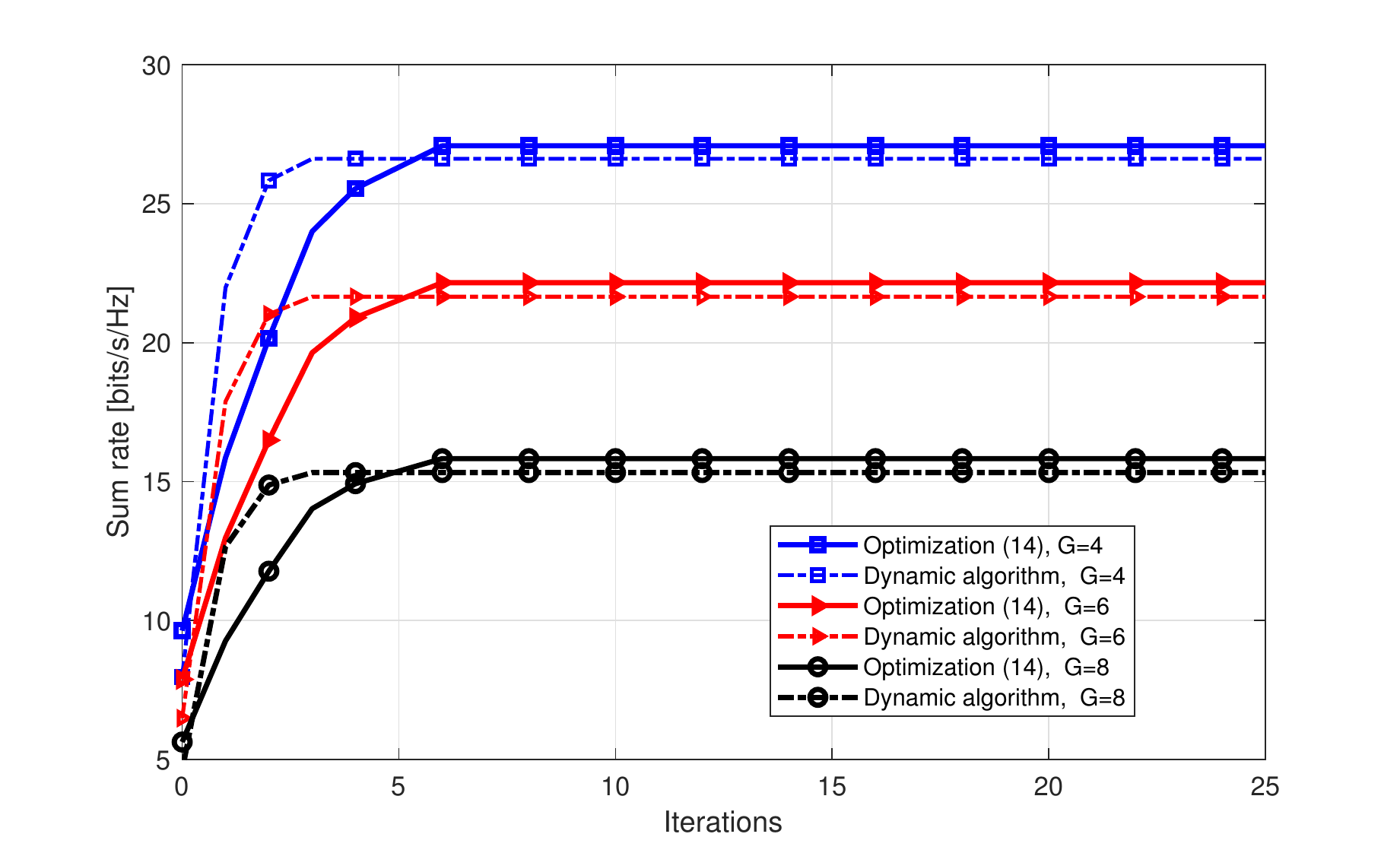}
\end{center}
\vspace{-2mm}
\caption{User grouping versus a set of iterations for the proposed dynamic algorithm.}\label{RS_fig1}
\vspace{-2mm}
\end{figure}

In Fig.~\ref{RS_fig1}, the sum rates of the user grouping methods are depicted against the iterations considering different numbers of groups $ G=\{ 4, 6, 8\} $. It is shown that the dynamic algorithm provides sub-optimal solutions significantly close to the optimal solutions obtained from solving the optimization problem in  \eqref{op2} through exhaustive search. It is worth noticing that the number of groups affects the sum rate of the proposed limited CSI-RS scheme with fixed power allocation. For instance, the proposed scheme provides higher sum rate 26.6 [bits/s/Hz] with $ G=4 $, i.e., $ \mathcal{K}_{m}=4 $ and $ \mathcal{K}_{ed}=16 $, compared to the scenarios of $ 6 $ and $ 8 $ groups. This is because of the trade-offs between the sum rate and the number of groups. In other words, increasing the number of groups results in a larger transmission block and noise enhancement (see equations \eqref{ratec}, \eqref{noisRS} and \eqref{ratep}), which limit the performance of BIA used as an outer precoder to align inter-group interference. Note that, having a high number of groups reduces the complexity of MUI management belonging to each group using RS. However, it is at the cost  of enhancing the noise among the formed groups.

\begin{figure}[t]
\begin{center}\hspace*{0cm}
\includegraphics[width=0.8\linewidth]{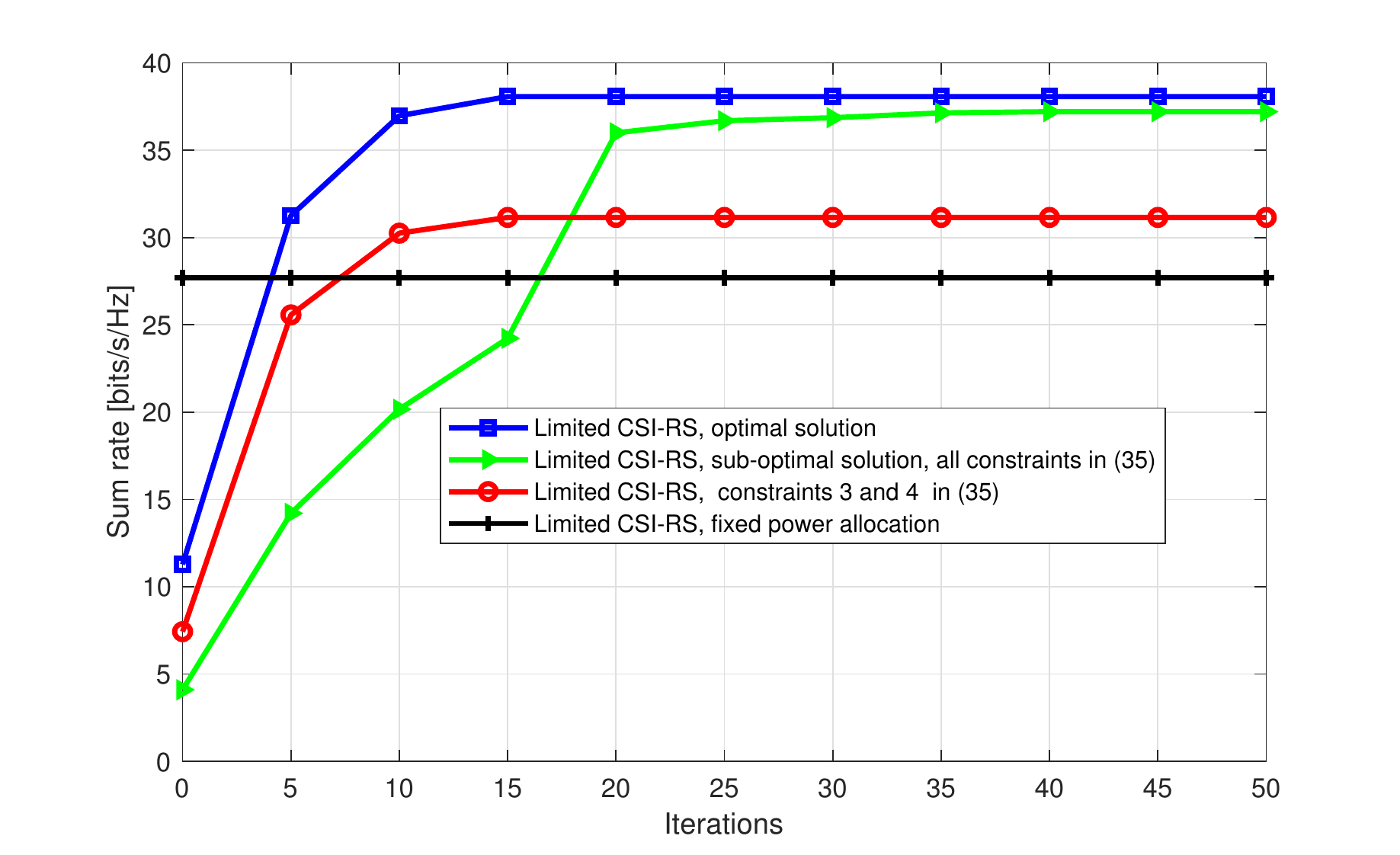}
\end{center}
\vspace{-2mm}
\caption{The sum rate of the proposed limited CSI-RS scheme after power allocation under different constraints. $ K=20 $, $ \mathcal{K}_{m}=4 $ and $ \mathcal{K}_{ed}=16 $. }\label{RS_fig2}
\vspace{-2mm}
\end{figure}

In Fig.~\ref{RS_fig2}, the performance of the proposed scheme is shown against a set of iterations after performing the optimization of power allocation. It can be seen that the sum rate of the proposed scheme can be enhanced considerably after maximizing the minimum sum rate of the users belonging to each group where the power budget can be utilized efficiently compared to simply dividing the power among the common and private messages  intended to the users regardless of their requirements. The figure further shows that the reformulation of the optimization problem using multiple multipliers relaxes the complexity while providing solutions close to the optimal. Specifically, the optimization problem is solved using the multipliers $ \beta^{[k]}_{g} $ and $ \nu_{g} $ in the red curve, and therefore, it can be seen that the sum rate is 30.2 [bits/s/Hz] compared to 38 [bits/s/Hz] achieved from solving the main power allocation problem, which is due to the fact that the power is allocated to maximize the sum rate of each group independently and without guaranteeing the quality of service of each user, i.e., without the constraints  $ \min\limits_{g \in G}  \Big\{ {R^{[g]}_{\text{sum}}} (P^{\dagger}_{c},P^{[k,g]}_{p})- \rho  \Big\} \leq {R^{[g]}_{\text{sum}}} (P^{\dagger}_{c},P^{[k,g]}_{p})- \rho
$ and $ {R^{[g]}_{\text{sum}}} (P^{\dagger}_{c},P^{[k,g]}_{p}) \geq R^{[g]}_{min} $. On the other hand, when the overall algorithm iterates over the four multipliers $ \beta^{[k]}_{g} $, $ \nu_{g} $, $ \lambda_{g} $ and $ \xi_{g}$ in the green curve, a significant solution is provided close to the optimal one. This solution is considered for the rest of the results due  to its practicality in terms of complexity.

\begin{figure}[t]
\begin{center}\hspace*{0cm}
\includegraphics[width=0.8\linewidth]{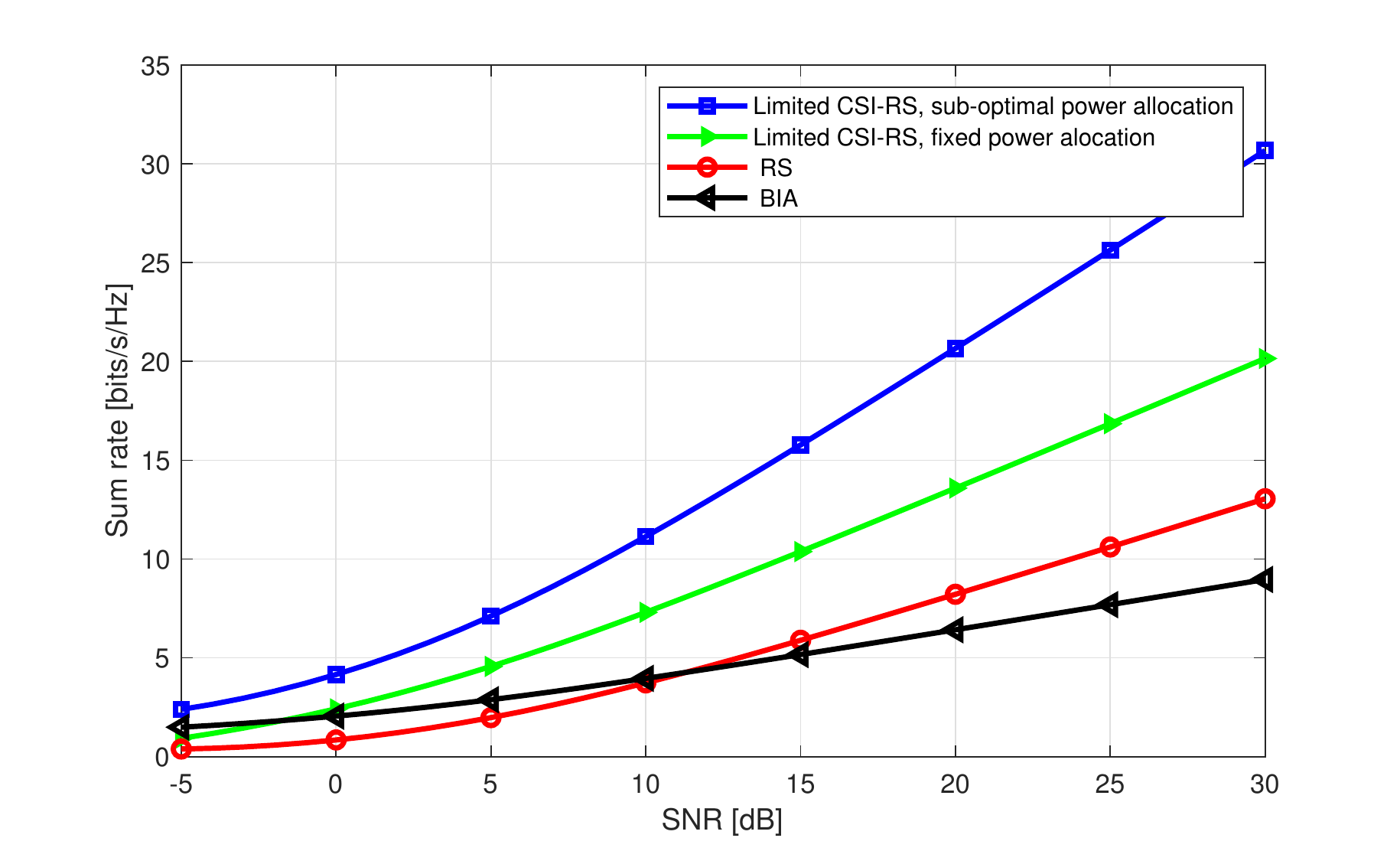}
\end{center}
\vspace{-2mm}
\caption{The sum rate of the limited CSI-RS scheme versus SNR compared with benchmark schemes. $ K=20 $, $ \mathcal{K}_{m}=4 $ and $ \mathcal{K}_{ed}=16 $.}\label{RS_fig3_2}
\vspace{-2mm}
\end{figure}

The sum rate of the proposed scheme  is shown in Fig.~\ref{RS_fig3_2} against a range of SNR values to demonstrate its superiority compared with BIA and RS schemes used to derive  its principles. It is shown that the proposed scheme provides higher sum rates than BIA and RS at different SNR values. It is worth mentioning that   the proposed scheme divides the users into multiple groups, and then, by following the transmission block of BIA determined by the number of groups, the outer precoding matrices of the groups are designed with no CSI to transmit the common and  private messages intended to each group. In BIA, the number of users determines the length of the transmission block, and each user must subtract the information transmitted to $ K-1 $ users. Therefore, the sum rate of the users slightly increases with increase in the SNR. Furthermore, RS suffers performance degradation in a high density network consisting of a high number of transmitters  serving  multiple users due  to the cost  of providing  CSI at transmitters in such scenarios, and the fact that each user performs SIC to cancel  the interference caused by the transmission to all other users, which has a negative impact on the achievable sum rate of the network.

\begin{figure}[t]
\begin{center}\hspace*{0cm}
\includegraphics[width=0.8\linewidth]{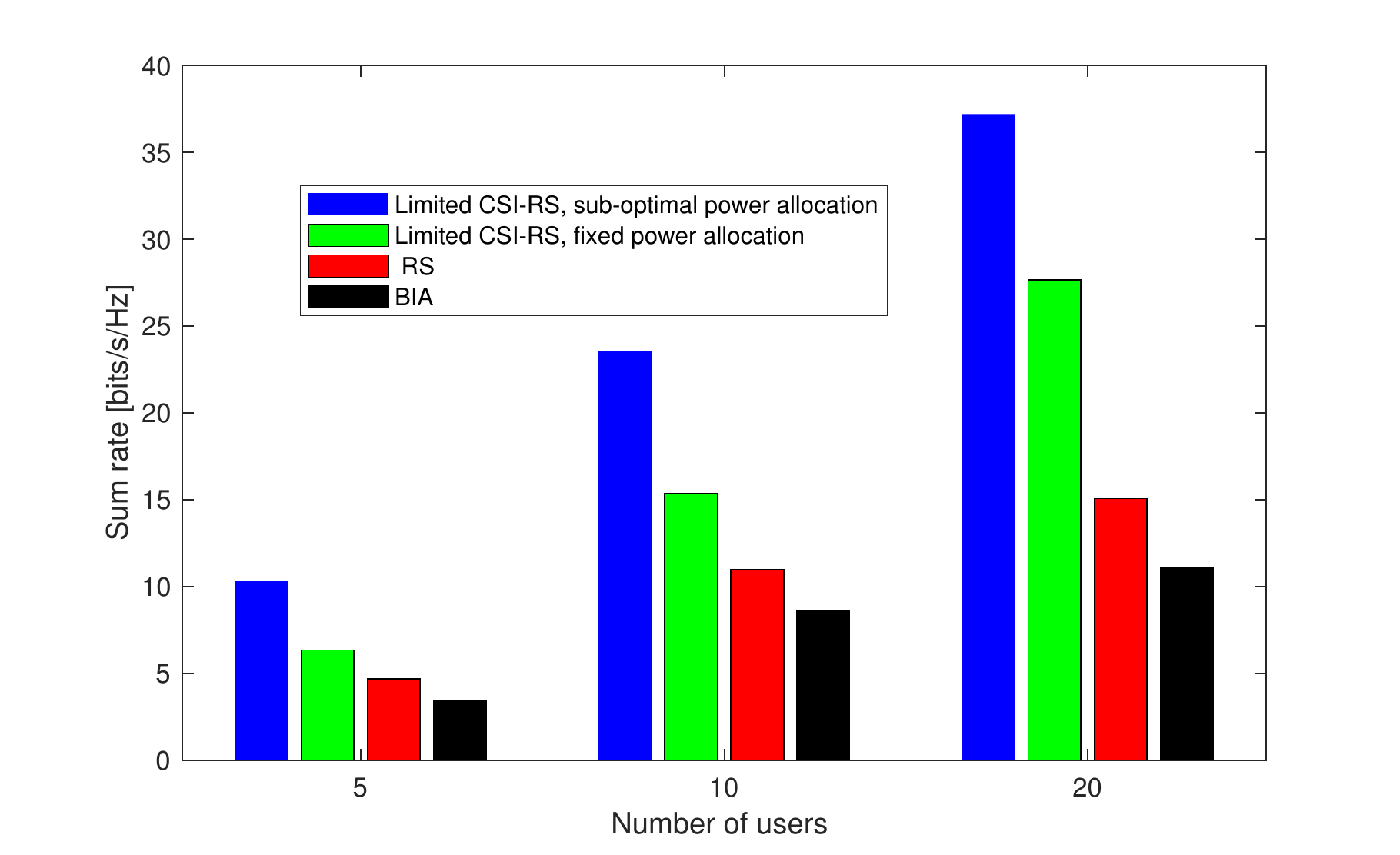}
\end{center}
\vspace{-2mm}
\caption{The sum rate of the limited CSI-RS scheme versus $ K $ users compared with benchmark schemes. }\label{RS_fig5}
\vspace{-2mm}
\end{figure}

In Fig.~\ref{RS_fig5}, the performance of the proposed scheme is depicted against different numbers of users  compared to BIA and RS. It can been seen that the sum rate of the network increases with the number of users regardless of the  transmission scheme considered for multi-user interference management  since it represents the aggregate data rate determined from the sum of the user rates. The proposed scheme is more suitable for the OWC network as the number of users increases, providing higher sum rate in all the scenarios  due  to its unique way in  handling interference while minimizing the noise resulting from interference cancellation. Note that, the application of the proposed power allocation technique further enhances the performance of the limited CSI-RS scheme compared to the fixed power allocation method. Moreover, the sum rates of RS and BIA schemes  increase slightly with the number of users as each user receives a  low data rate subject to high noise and errors caused by the limitations of these schemes in serving such a high number of users.

\begin{figure}[t]
\begin{center}\hspace*{0cm}
\includegraphics[width=0.8\linewidth]{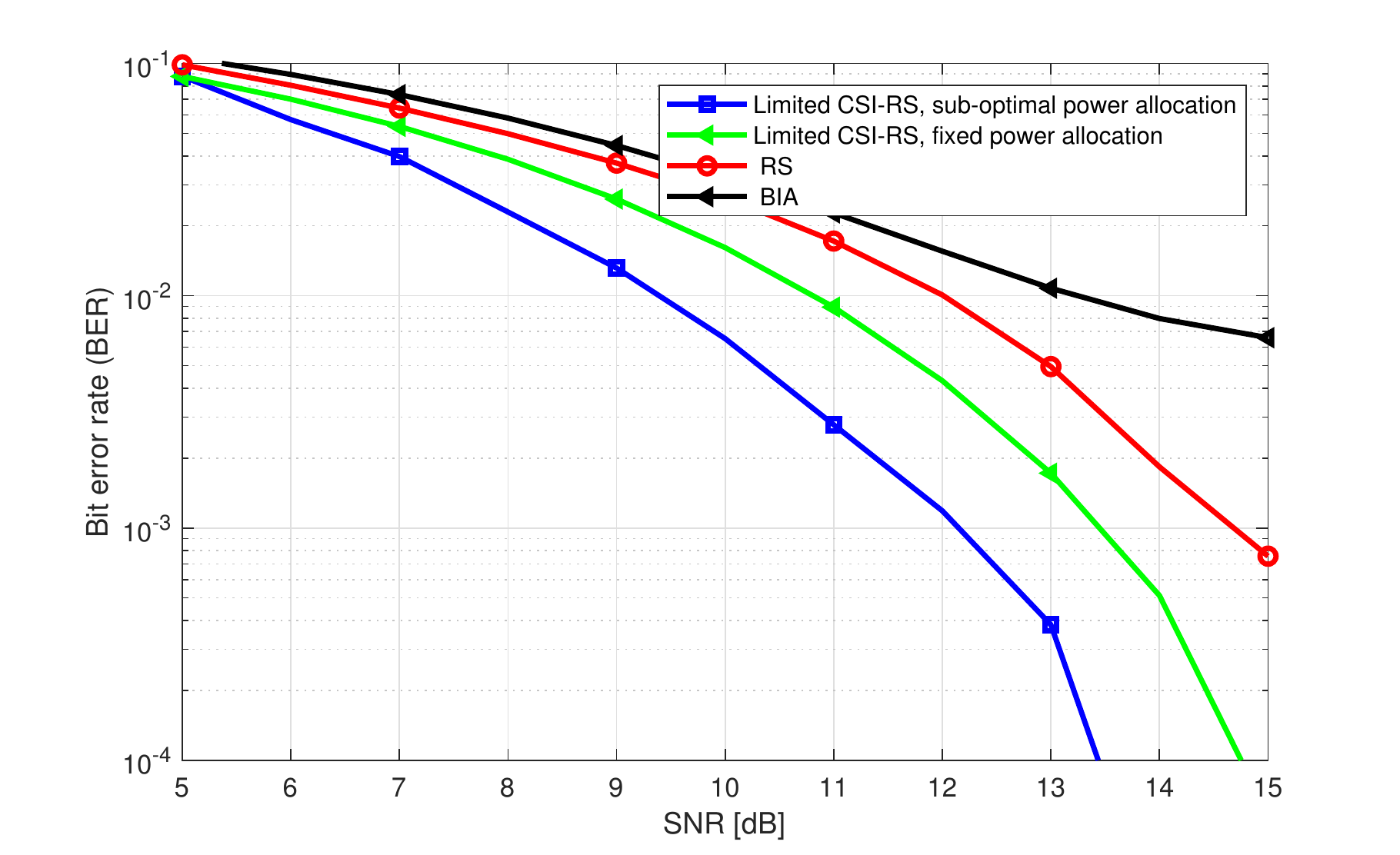}
\end{center}
\vspace{-2mm}
\caption{BER for 2-PAM Modulation. $ \mathcal{K}_{m}=4 $ and $ \mathcal{K}_{ed}=16 $.}\label{RS_fig4}
\vspace{-2mm}
\end{figure}

To evaluate the BER of the limited CSI-RS scheme, multilevel N-ary pulse amplitude modulation (N-PAM) is considered  where the mean of the transmitted signal is equal to the current, which ensures the linear response of the optical transmitter. In Fig.~\ref{RS_fig4}, BER  for $ K=20 $ users is plotted against a range  of  SNR values, (5,15) dB, considering  different transmission schemes. It can be seen that the limited CSI-RS is superior compared to RS and BIA schemes at different SNR values. For instance, the proposed scheme achieves a BER value less than $ 10^{-3} $ at 14 dB SNR with and without the power allocation optimization problem, while RS and BIA schemes at the same value of SNR achieve BER values higher than  $ 10^{-3} $  and slightly less than  $ 10^{-2} $, respectively. This is owing to the fact that each user $ k $ experiences high noise resulting from the interference cancellation of the information transmitted to all other users, $ k'\neq k $, in the case of RS and BIA schemes.

\begin{figure}[t]
\begin{center}\hspace*{0cm}
\includegraphics[width=0.8\linewidth]{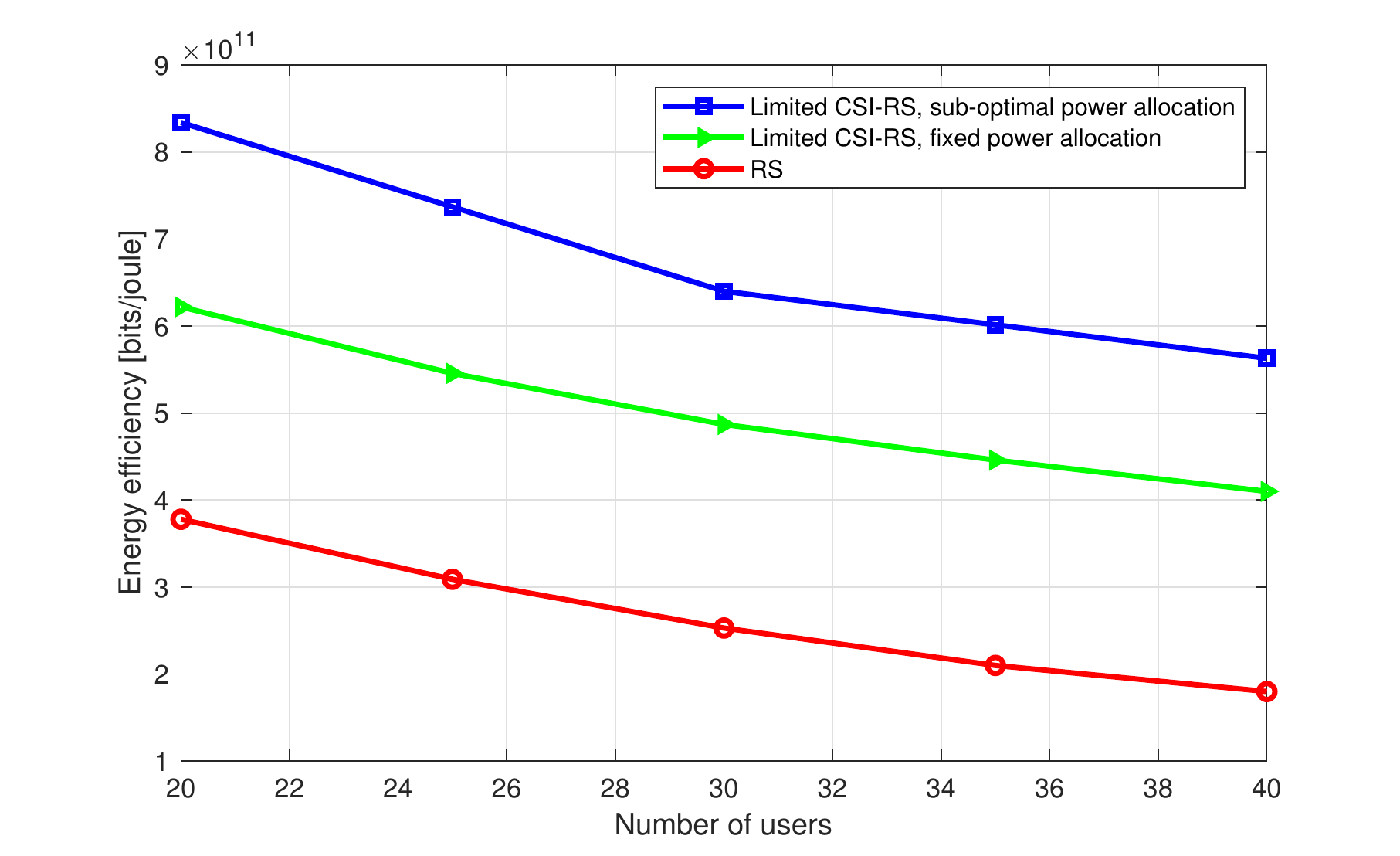}
\end{center}
\vspace{-2mm}
\caption{Energy efficiency of the limited CSI-RS scheme versus $ K $ users.}\label{RS_fig6}
\vspace{-2mm}
\end{figure}

To further test the effectiveness of the proposed scheme, its energy efficiency, which is one of the most important metrics in wireless communications, is depicted in Fig.~\ref{RS_fig6} against a higher number of users compared with the conventional RS scheme. It can be seen that the energy efficiency of the network decreases as the number of users increases up to $ 40 $ users due to the fact that  more power is consumed at a high number of users. However, the limited CSI-RS scheme is more energy efficient than the RS scheme as the number of users increases from 20 to 40 users. This behavior is expected due to the implementation of the user grouping algorithm and the use of the BIA-based outer precoder to eliminate inter-group interference. Furthermore, it can be seen that the overall power allocation algorithm distributes the power among the messages of the users more efficiently based on the sum rate maximization of the network resulting in higher energy efficiency compared with the fixed power allocation scheme that has low complexity at the cost of serving users regardless of the quality of service.   

\section{Conclusions}
\label{sec:con}
In this work, a novel RS-based transmission scheme is proposed in an OWC network to serve multiple users simultaneously maximizing the spectral efficiency of  the network. We first formulate an optimization problem where users are divided into multiple groups under fixed power allocation. Then, a dynamic algorithm is designed to relax the complexity, while guaranteeing the formation of each group with a unique set of users. After that, a limited CSI-RS scheme is derived to manage inter-group interference using an outer precoder designed with limited CSI to the channel coherence time and the distribution of users, while the users belonging to each group receive their desired information following the methodology of RS. Finally, an optimization problem is formulated to maximize the minimum sum rate within each group through finding the optimum power allocated to the private message intended to each user. The power allocation is reformulated via multiple multipliers with the aim of reducing complexity, while providing sub-optimal solutions. The results show that the proposed limited CSI-RS scheme achieves high performance in terms of sum rate, BER and energy efficiency compared to the benchmark schemes considered.

%\begin{thebibliography}{1}
       % \bibitem{IA} V.R. Cadambe, and S.A. Jafar,
      %  ``Interference Alignment and Degrees of Freedom of the  K -User Interference Channel,''
     %   {\it IEEE Trans. on Information Theory}, vol. 46, no. 9, pp. 59-67, September 2008.
%\bibitem{I} A.R. Cadambe, and S.A. Jafar,
        %``Interference Alignment and Degrees of Freedom of the  K -User Interference Channel,''
       % {\it IEEE Trans. on Information Theory}, vol. 46, no. 9, pp. 59-67, September 2008.
%\end{thebibliography}

%\bibliographystyle{IEEEtran}

%\bibliography{IEEEabrv,mybib}

%\bibliographystyle{IEEEtran}

%\bibliography{IEEEabrv,ahbib}
\bibliographystyle{IEEEtran}
\bibliography{IEEEabrv,mybib}

\end{document}